%% file: root.tex
\documentclass[twocolumn]{autart}
\include{preamble_Automatica}

\begin{document}

\begin{frontmatter}
\title{Data-Driven Abstractions for Control Systems via Random Exploration}

\thanks[footnoteinfo]{R. Coppola, A. Peruffo and M. Mazo Jr. are with Faculty of Mechanical Engineering,
        TU Delft, Delft, The Netherlands.
        {\tt\small r.coppola@tudelft.nl} }

\author[TUD]{Rudi Coppola}\ead{r.coppola@tudelft.nl},    
\author[TUD]{Andrea Peruffo}\ead{a.peruffo@tudelft.nl},               
\author[TUD]{Manuel Mazo Jr.}\ead{m.mazo@tudelft.nl}  

\address[TUD]{Technical University of Delft, Delft}  


\begin{abstract}                          
At the intersection of dynamical systems, control theory, and formal methods lies the construction of symbolic abstractions: these typically represent simpler, finite-state models whose behavior mimics that of an underlying concrete system but are easier to analyse. 
Building an abstraction usually requires an accurate knowledge of the underlying model: this knowledge may be costly to gather, especially in real-life applications. 
We aim to bridge this gap by building abstractions based on sampling finite length trajectories. 
To refine a controller built for the abstraction to one for the concrete system, 
we newly define a notion of probabilistic alternating simulation, and provide Probably Approximately Correct (PAC) guarantees that the constructed abstraction includes all behaviors of the concrete system and that it is suitable for control design, for arbitrarily long time horizons, leveraging scenario theory. 
Our method is then tested on several numerical benchmarks.
\end{abstract}

\end{frontmatter}

\section{Introduction}
\label{sec:intro}

Data-driven modeling and analysis is undergoing a new renaissance with the advances in machine learning and artificial intelligence enabled by unprecedented computing power. The field of system verification, aimed at providing formal performance and safety guarantees is not alien to this trend. Recent work has been focused on the use of collected data from a system to derive directly (i.e. with no model involved) barrier functions certifying invariance~\cite{akella2022barrier,salamati2021data}, or finite abstractions to verify and synthesize controllers~\cite{devonport2021symbolic,majumdar2020abstraction,sadraddini2018formal}. A popular approach is to employ scenario-based optimization techniques to derive probably approximately correct (PAC) guarantees on the performance metric of interest.
The use of scenario-based optimization 
requires independent samples, generated from the probability distribution that drives the system's uncertainty.
A special case considered in many works \cite{devonport2021symbolic,wang2020scenario} is that of deterministic systems (for which a model is not available), in which the only uncertainty is their initialization, i.e. the initial state is drawn from some probability distribution, and 
the control policy, usually chosen from a finite set of possible actions, and the samples consist of individual transitions or, more generally, finite-length trajectories.
In this setting the independence of the samples can be derived by independently sampling the initial state;  typically a uniform distribution is selected for compact initial sets. 
%
If the set of initial conditions coincides with the domain of interest, one can directly sample one-step transitions. However, the scenario-based approach would ensure the satisfaction of \emph{one-step} properties, 
provided that transitions are indeed sampled independently.  
On the other hand, one is often interested in inferring long (even infinite) horizon specifications from one-step properties. These approaches are limited to provide guarantees for the time horizon used to construct the data set, as for larger horizons nothing can be directly inferred unless some additional knowledge of the system is available, as for example in \cite{lavaei2022constructing}. Addressing this limitation in the context of abstraction-based control synthesis is the main objective of this work. 
\newline
\textbf{Contributions.}
We consider deterministic control systems with unknown dynamics and a random initialization. 
Our approach provides a construction of data-driven finite abstractions, built on the notions of alternating simulation and a particular class of abstractions known as Strongest Asynchronous $\ell$-complete Abstractions (or Approximations) (SA$\ell$CAs) \cite{schmuck2015comparing}. Unlike our previous work \cite{coppola2022data,coppola2023data}, which treats verification problems, the data-driven abstraction presented in this paper enables the synthesis of a controller for the unknown system to solve reach-avoid specifications.
We introduce a notion of probabilistic alternating simulation, instrumental in describing the relation between a deterministic (but randomly sampled) model and a transition system constructed upon the collected system's behaviors. 
Leveraging the scenario theory, we establish PAC guarantees for the inclusion of the concrete system's finite behaviors in those of the abstraction. We clarify the role of the trajectory length used for constructing the abstraction; successively, if the system's dynamics is \emph{partially known}, we propose sufficient conditions to extend PAC guarantees over \emph{longer} trajectory lengths. 
This enables the verification of properties, and synthesis of control policies, over an arbitrarily long time horizon while preserving the PAC guarantees for several classes of nonlinear systems.
\newline
\textbf{Related Work.} 
Abstractions simplify complex systems by reducing them to finite-state models, aiding in property verification and controller synthesis \cite{tabuada2009verification}. 
Recent methods bypass the need for an explicit model by directly synthesizing abstractions from data. 
In \cite{cubuktepe2020scenario,badings2023robust,lavaei2022constructing}, a sampled-based interval Markov Decision Process is created using the scenario approach to constrain transition probabilities in a stochastic dynamical model. 
In \cite{makdesi2023data}, PAC over-approximations of monotone systems are employed to construct models; in \cite{kazemi2022data}, a sample-based growth rate is used to build an abstraction and synthesize a controller. 
In \cite{devonport2021symbolic}, the authors pursue a goal, similar to ours, of deriving an  abstraction from a black-box system suitable for control, and propose a notion of PAC (approximate) alternating simulation relationship. Unfortunately, the authors' procedure to derive an abstraction relies on chaining one-step transitions sampled uniformly: this violates the PAC bound assumptions that the authors adopt, thus the abstraction does not satisfy the proposed notion, see \cite[Section 3.3]{coppola2022data}.
\newline
In contrast, \cite{ajeleye2023data} efficiently constructs an abstraction of a deterministic control system with disturbances by estimating reachable sets, assuming the Lipschitz constants are known.
In the first part of this work, we show how to derive an abstraction suitable for control for a black-box system for a finite horizon, without assuming any limiting knowledge on the dynamics, in contrast to \cite{ajeleye2023data}. Only later, for the purpose of extending this horizon, we introduce additional assumptions, involving for instance the Lipschitz constant.
\newline
%
%
%
Our work focuses on a specific type of abstraction known as SA$\ell$CA \cite{schmuck2014asynchronous,schmuck2015comparing}, which offers several useful properties for our purposes. Unlike state-based abstractions, a SA$\ell$CA can be constructed from complete knowledge of a system's output trajectories. However, when only partial knowledge is available, such as when using sampled trajectories, certain challenges arise, as detailed in \cite{coppola2022data}.
In \cite{coppola2023data}, the authors extend SA$\ell$CAs to linear autonomous systems using data-driven approaches. Our work builds on this by extending these methods to nonlinear control systems. SA$\ell$CAs have been applied in event-triggered control models \cite{peruffo2022data,peruffo2023sampling}.
Recent research also explored data-driven memory-based Markov models for stochastic systems \cite{banse2023data,banse2023adaptive}, emphasizing the importance of memory in constructing effective abstractions. As we show, memory plays a crucial role in leveraging data-driven SA$\ell$CAs for control applications.
\newline
\textbf{Organisation. }
Section \ref{sec:preliminaries} provides relevant background information. 
Section \ref{sec:problem} describes transition systems and system relations, the primary tools for deriving an abstraction-based controller. 
Section \ref{sec:data-driven} introduces a data-driven approach to constructing abstractions for \emph{unknown} (black-box) systems, using scenario theory to establish PAC-type guarantees for the horizon used for sampling; Section \ref{sec:greater-horizons} extends these guarantees to longer time horizons by studying \emph{partially known} systems, making specific assumptions about the dynamics. 
Section \ref{sec:experiments} and Sections \ref{sec:discussion} are dedicated to showcase experimental results, a discussion and a conclusion.


\section{Notation and Preliminaries}\label{sec:preliminaries}


Given a set $\mathcal{M}$, we denote its $n$-th cartesian product by $\mathcal{M}^n$, and its power set by $\wp(\mathcal{M})$. Let $s=m_0a_0...a_{n-1}m_n$ be any sequence such that $m_i$'s belong to a set $\mathcal{M}$ and $a_i$'s belong to a set $\mathcal{A}$. We denote $s(i)$ the $i$-th element belonging to $\mathcal{M}$, i.e. $s(i)=m_i$, for $i=0,...,n$; by $s[i,i+j]$, with $j\geq0$, we denote the $j$-long subsequence of $s$ from $m_i$ to $m_{i+j}$, i.e. $s[i,i+j]=m_ia_i...a_{i+j-1}m_{i+j}$. Given two sequences $s=m_0a_0...a_{i-1}m_i$ and $s'=m_0'a_0'...a_{j-1}'m_j'$ with $m_i=m_0'$, we denote their concatenation by $s\cdot s' \doteq m_0a_0...a_{i-1}m_0'a_0'...a_{j-1}'m_j'$. Finally, $s|^m$ ($s|^a$) denotes the sequence obtained by removing from $s$ all the elements that do not belong to $\mathcal{M}$ ($\mathcal{A}$), and $s|^m(i)=m_i$.
\newline
For a relation $\cR\subseteq \cX_a\times\cX_b$, we define $\cR_{\cX_b}(x_a) \doteq \{x_b\in\cX_b: (x_a,x_b)\in\cR\}$. We indicate the inverse relation of $\cR$ by $\cR^{-1}$, i.e. $(x_b,x_a)\in\cR^{-1} \iff (x_a,x_b)\in\cR$.
\newline
Denote by $(\Omega_i, \cF_i, \mu_{\omega_i})$ for $i=1,2$, two probability spaces, where $\Omega_i$ is the sample space, endowed with a $\sigma$-algebra $\mathcal{F}_i$ and a probability measure $\mu_{\omega_i}$. We define the product probability space $(\cP,\mathcal{W},\mu_p)$ where $\cP = \Omega_1\times \Omega_2$,
denoted $\cW=\cF_1\times\cF_2$ is the product $\sigma$-algebra , and $\mu_p=\mu_{\omega_1} \times \mu_{\omega_2}$ is the product measure.


\subsection{Dynamical Models}
We consider a time-invariant dynamical system 
\begin{equation}
\label{eq:deterministic-sys}
   \Sigma(x) \doteq 
   \begin{cases}
       x_{k+1} = f( x_k, u_k ), 
       \\ 
       y_k = h(x_k),
       \\
       x_0 = x,
       \end{cases}
\end{equation}
where $x_k \in \cX \subset \reals^{n_x}$ is the system’s state at time $k \in  \naturals_0$ (natural numbers including zero), $n_x$ is the state-space dimension, $x_0$ is the initial state,  $y_k \in \cY$ is the system output where $\cY$ is an arbitrary output set with $|\cY|<\infty$, $u_k\in \cU \subset\reals^{n_u}$ is the system input at time $k$, $\cU$ is a finite input set, i.e. $|\cU|<\infty$, $n_u$ is the input dimension. 
We denote as $\mathbf{u}_H\in\cU^H$ a sequence of control inputs of length $H$. We assume that $f(\cdot,u)$ is measurable on the standard Borel space associated with $\mathbb{R}^n$ for all $u$.

\begin{definition}
\label{def:Lipschitz-inv}
    Let $(\cX,d_X)$ and $(\cU,d_U)$ be a complete metric spaces. Let the map $f$ defined in \eqref{eq:deterministic-sys} satisfy the inequality $
        m_Xd_X(x,x')\leq d_X(f(x,u),f(x',u))$,
    for $m_X>0$ all $x\neq x'\in\cX$ and $u\in\cU$. Then, $f$ is said to be Lipschitz invertible.
    \newline
    The map $f$ in \eqref{eq:deterministic-sys} is uniformly contracting w.r.t. $x$ if there 
exists $0< l_X < 1$ such that $
d_X( f(x,u), f(x',u) ) \leq l_X d_X(x,x'),
$,
for all $x,x' \in \mathcal{X}, \text{ and } u \in \mathcal{U}.$
Similarly, $f$ is uniformly Lipschitz w.r.t. $u$ if there exists $l_U > 0$ such that
$
d_X ( f(x,u),f(x,v))  \leq l_U d_U(u,v),
$,
$\text{ for all } x\in \mathcal{X}, \text{ and } u,v \in \mathcal{U}.$ 
 
\end{definition}


\subsection{Scenario Theory}
\label{subsec:scenario}

Let $(\Omega,\mathcal{F},\mu)$ be a probability space, and consider $N$ independent $\mu$-distributed samples $(\omega_1 , \ldots , \omega_N )$.
Each $\omega_i$ is regarded as an observation, or \emph{scenario} \cite{campi2018general,garatti2021risk}.
We aim at taking a \emph{decision}, $\theta^*_N$, e.g. construct a classifier, from a set  $\Theta$,  the decision space.  To every $\omega \in \Omega$ there is associated a constraint set $\Theta_\omega \subseteq \Theta$.
The scenario theory is a \emph{distribution-free} setting that, under very mild assumptions \cite[Assumption 1]{garatti2021risk}, offers a way to compute a decision $\theta_N^*$ satisfying all the constraints $\Theta_{\omega_i}$ imposed by the set of $N$ i.i.d. samples while quantifying the probability that a new random sample $\omega\in\Omega$ would result in the solution $\theta^*_N$ violating the new constraint $\Theta_{\omega}$.
Hence, it quantifies the generalization power of the solution.
\begin{theorem}
\label{theo:scenario-gurantees}
\textnormal{\textbf{(PAC bounds {\cite[Theorem~1]{garatti2021risk}})}}  For a given $\theta \in \Theta$ let $\mathcal{V(\theta)} = \{\omega : \theta\notin\Theta_{\omega}\}$ be the \emph{violation set} and $V ( \theta ) = 
    \mu ( \, \omega \in \mathcal{V(\theta)})$ be the \emph{violation probability} (or violation for short).  
For a confidence $\beta \in (0, 1)$ and decision $\theta^*_N$, it holds
\begin{equation}
\label{eq:scenario-confidence}
    \mu^N (
    V(\theta^*_N) \leq \epsilon(s^*_N, \beta, N)
    ) \geq 1 - \beta,
\end{equation}
 where $\epsilon(\cdot)$ is the solution of a polynomial equation (omitted here for brevity) and $s^*_N$ is the \emph{complexity} of the solution --  it represents the cardinality of the smallest subset of the samples yielding the same solution $\theta^*_N$. 
\end{theorem}
\begin{remark}\label{rem:degenerate-problem}
In this work we consider a discrete sample space $\Omega$, therefore we refer to the scenario theory for degenerate problems, as per \cite{garatti2021risk,campi2018general}.
\end{remark}


\section{Problem Statement and System Description}
\label{sec:problem}

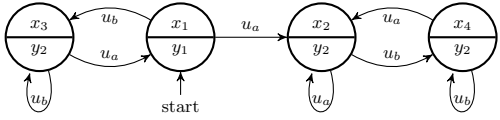
\begin{figure}[t]
    \centering
    \input{models/sm-example}
    \caption{Illustration of the TS of Example~\ref{example:TS}.}
    \label{fig:state-machine}
\end{figure}

We adopt the framework of finite-state abstractions in the form of transition systems.
\begin{definition}\label{def:state-machine}
    \textnormal{\textbf{(Transition System (TS))}} A transition system $S$ is a tuple $(\cX,\cX_0,\cU,\delta,\cY,\cH)$, where $\cX$ is the set of states, $\cX_0\subseteq\cX$ is the set of initial states, $\cU$ is the input set, $\cY$ is the set of outputs, $\delta\subseteq \cX\times\cU\times\cX$ is a transition relation, and $\cH:\cX\rightarrow \cY$ is an output map. 
     
\end{definition}
\vspace{-.2cm}
We define the set of $u$-successor states of a state $x$ as $\post{u}{x} \doteq {x' \in \cX : (x,u,x') \in \delta}$ and the set of admissible inputs at $x$ as $U_{\delta}(x) \doteq {u \in \cU : \post{u}{x} \neq \emptyset}$. If $U_{\delta}(x) \neq \emptyset$ for all $x$, the system is \emph{non-blocking}.
An $H$-long internal behavior of the TS, $\xi = x_0u_0x_1\ldots u_{H-1}x_H$, satisfies $x_0 \in \cX_0$ and $(x_{i-1},u_{i-1},x_i) \in \delta$ for all $i = 1, \ldots, H$. An $H$-long external behavior, $\gamma = y_0u_1y_1 \ldots u_{H-1}y_H$, satisfies $y_i = \cH(x_i)$ for all $i = 0, \ldots, H$. We denote the external behavior $\gamma$ corresponding to the internal behavior $\xi$ as $\cH(\xi)$. The sets $\cI_H(S)$ and $\cB_H(S)$ contain all $H$-long internal and external behaviors of the TS, respectively.
\newline
Given $x\in\cX_0$ and an $H$-long input sequence $\mathbf{u}_H\in\cU^H$ we define respectively the set of internal and external $H$-behavior of a TS $S$ starting in $x_0$ under input sequence $\mathbf{u}_H$ as
\vspace{-.2cm}
\begin{multline}
\label{eq:behavior}
    \cI_H(S,x_0,\mathbf{u}_H) \doteq  \{\xi \in \cI_H(S) : \\
    \xi(0) = x_0 \ \wedge
    \xi|^u=\mathbf{u}_H\},
\end{multline}
\vspace{-1cm}
\begin{multline}
    \label{eq:external-behave}
    \cB_H(S,x_0,\mathbf{u}_H)  \doteq  \{\gamma\in\cB_H(S) :  \\
     \exists \xi\in\cI_H(S,x_0,\mathbf{u}_H) \ . \ \gamma=\cH(\xi)\}.
\end{multline}
\vspace{-.5cm}
\newline
Our technique relies on the concept of memory via \emph{$\ell$-sequences}, which are $\ell$-long subsequences of external behaviors. Each state $x \in \cX$ is linked to all possible $\ell$-sequences that could lead to it, representing the recent past in terms of inputs and outputs. For initial states $x \in \cX_0$, following \cite{schmuck2015comparing}, we extend behaviors to negative time indices using the symbol $\diamond$, so $\xi|^x(k) = \xi|^u(k) = \gamma|^y(k) = \gamma|^u(k) = \diamond$ for $k \leq -1$.
\newline
Consider the set of external behaviors of length $H$ and an integer $\ell$ with $0 \leq \ell < H$. Each $\gamma \in \cB_H(S)$ is divided into $\ell$-long subsequences (or $\ell$-sequences). The set of all $\ell$-sequences from these behaviors is denoted by 
\vspace{-.2cm} 
\begin{equation}\label{eq:domino-tiles}
\Pi_{\ell,H} \doteq \bigcup_{\gamma\in\cB_H(S)}\bigcup_{k\in[0,H]}\gamma [k-\ell,k]. , \end{equation} 
\vspace{-.4cm} 
\newline Each $\zeta \in \Pi_{\ell,H}$ contains $\ell+1$ outputs and $\ell$ inputs. We omit the dependence of $\Pi_{\ell,H}$ on $S$ since it always refers to the concrete system.
The set of corresponding external strings (CESs) of length $\ell$ for a state $x \in \cX$ is defined as 
\vspace{-.2cm} 
\begin{multline}\label{eq:cor-ext-beh}
\cE_{\ell,H}(x)  \doteq  \{\zeta\in\Pi_{\ell,H} \ : 
    \exists \xi\in\cI_H(S),\\ 
    \exists j \in \naturals_0  \ . \ \zeta = \cH(\xi[j-\ell,j])
   \wedge \xi(j) = x\}.
\end{multline} 
$\cE_{\ell,H}(x)$ represents all subsequences of external behaviors with $\ell$ outputs that the system can generate before reaching $x$ in at most $H$ steps. If $x$ is reached in fewer than $\ell$ transitions from an initial state, some CESs will include the symbol $\diamond$. For instance, if $\ell=3$ and $x'$ is reached in one transition from the initial state $x$ by choosing control input $u$, then $\diamond\diamond\cH(x)u\cH(x')\in\cE_{3,H}(x')$.
The equivalence class of an $\ell$-sequence $\zeta\in\Pi_{\ell,H}$ is 
\vspace{-.1cm}
\begin{equation}\label{eq:eq-classes}
    [\zeta]  \doteq  \{x :\zeta\in\cE_{\ell,H}(x)\}.
\end{equation}
\vspace{-.5cm}
\begin{example}\label{example:TS} 
Let $S$ be the TS depicted in Fig.~\ref{fig:state-machine}, where $\cX = \{x_1, x_2, x_3, x_4\}$, $\cX_0 = \{x_1\}$, $\cU = \{u_a,u_b\}$, and $\cY=\{y_1,y_2\}$. 
For a time horizon $H=4$ and input sequence $\mathbf{u}_3 = u_a u_b u_a$. The internal and external behaviors initialized from $x_1$, are given by $
    \cI_3(S, x_1, \mathbf{u}_3) = \{ x_1 u_a x_2 u_b x_4 u_a x_2 \}$ and $
    \cB_3(S, x_1, \mathbf{u}_3) = \{ y_1 u_a y_2 u_b y_2 u_a y_2 \}$.
For $\ell=1$, following \eqref{eq:domino-tiles} we split $\cB_3(S, x_1, \mathbf{u}_3)\in\cB_H(S)$ in subsequences of length $1$ and conclude that $
    \{ y_1, y_2 \} = \cY
    = \Pi_{0,3}$.
Moreover, from \eqref{eq:cor-ext-beh} we have $y_1\in\cE_{1,3}(x_1)$, $y_2=\cE_{1,3}(x_2)$, and $y_2\in\cE_{1,3}(x_4)$.
For $\ell=2$, we have $
     \{ \diamond \diamond y_1, y_1 u_a y_2,
     y_2 u_b y_2, y_2 u_a y_2 \}\subset \Pi_{1,3}$.
 Moreover, $\diamond \diamond y_1\in\cE_{2,3}(x_1)$, $y_1 u_a y_2\in\cE_{2,3}(x_2)$, $y_2 u_b y_2\in\cE_{2,3}(x_4)$, and again $y_2 u_a y_2\in\cE_{2,3}(x_2)$.  
\end{example}

The system $\Sigma$ in \eqref{eq:deterministic-sys} can be equivalently described as a TS, potentially with an infinite state set, denoted by $S_{\Sigma}$. To account for transitions where $f(x,u) \notin \cX$ for some $x \in \cX$ and $u \in \cU$, we add an absorbing state $x_{\text{abs}}$ to the state set of $S_{\Sigma}$, where $f(x_{\text{abs}},u) = x_{\text{abs}}$ for all $u$ and with a unique output label $y_{\text{abs}}$. We assume $S_{\Sigma}$ satisfies Assumption \ref{ass:concrete-sys}.
\begin{assumption}\label{ass:concrete-sys}
   $S_\Sigma$ has \emph{free input}, i.e. $U_\delta(x) = \mathcal{U}$ for all $x$, is \emph{deterministic}: in short $|\post{u}{x}| = 1$ for every $x\in\cX$ and $u\in\cU$, and has $\cX_0=\cX$. 
\end{assumption}

\smallskip 

The assumption that $\cX_0=\cX$ simplifies our notation in that it implies that the horizon $H$ does not affect the definition of \eqref{eq:domino-tiles} and \eqref{eq:cor-ext-beh}, since every $\ell$-sequence that the system can generate at any point in time, can also be generated as an initial sequence. 
Formally, 
it holds that $\cE_{\ell,\ell+1}(x) = \cE_{\ell,H}(x)$ for every $H>\ell$. Moreover, $\bigcup_{x}\cE_{\ell,H}(x)=\Pi_{\ell,H}$. From now on, we drop the second subscript and denote the CESs strings of a state $x$ simply as $\cE_\ell(x)$ and the set of all $\ell$-sequences of all external behaviors as $\Pi_\ell$. The case with $\cX_0\neq\cX$ follows \emph{mutatis mutandis} with no conceptual modifications.

\subsection{Systems' Relations}
We recall two essential concepts: the \emph{alternating simulation relation} (ASR) and the \emph{simulation relation} (SR) \cite{tabuada2009verification,schmuck2015comparing}. The ASR is crucial for refining a policy derived from an abstraction into a controller for the concrete system. 
As in \cite{schmuck2015comparing}, we define an observation to include both input and output, unlike the more common definition where only the output is observed. This broader observation allows for constructing a more refined abstraction. For simplicity, we assume that the abstraction and the concrete system share the same input space.
\begin{definition}\label{def:sim-rel}
\textnormal{\textbf{(Simulation relation (SR) {\cite{schmuck2015comparing}})}} Consider two non-blocking systems $S_a$ and $S_b$ with $\cY_a=\cY_b$, and $\cU_a = \cU_b$. A relation $\cR\subseteq \cX_a\times\cX_b$ is a simulation relation from $S_a$ to $S_b$ w.r.t. $\cU\times\cY$, written $S_a  \preceq^{\cR}_{S.}  S_b$, if the following three conditions are satisfied:
\begin{itemize}
    \item $\forall x_{a0}\in\cX_{a0} \ . \ \exists x_{b0}\in\cX_{b0}$ with $(x_{a0},x_{b0})\in \cR$,
    \item $(x_{a},x_{b})\in \cR \implies \cH_a(x_a)=\cH_b(x_b)$,
    \item $(x_{a},x_{b})\in \cR \implies (U_{\delta_a}(x_a)\subseteq U_{\delta_b}(x_b)\wedge\forall u \in U_{\delta_a}(x_a) \ . \  
    (x_a,u,x_a')\in\delta_a \implies  \exists x_b'\in\cX_b \ . \ (x_b,u,x_b')\in\delta_b \wedge (x_a',x_b')\in\cR)$. 
\end{itemize} 
\end{definition}

\smallskip

\begin{definition}\label{def:alt-sim-rel}
\textnormal{\textbf{(Alternating simulation relation (ASR) {\cite{tabuada2009verification}})}} Consider two non-blocking systems $S_a$ and $S_b$ with $\cY_a=\cY_b$, and $\cU_a = \cU_b$. A relation $\cR\subseteq \cX_b\times\cX_a$ is an alternating simulation relation from $S_b$ to $S_a$ w.r.t. $\cU\times\cY$, written $S_b \preceq_{A.S.}^{\cR} S_a$, if the following three conditions are satisfied:
\begin{itemize}
    \item $\forall x_{b0}\in\cX_{b0} \ . \ \exists x_{a0}\in\cX_{a0}$ with $(x_{b0},x_{a0})\in \cR$,
    \item $(x_{b},x_{a})\in \cR \implies \cH_a(x_b)=\cH_b(x_a)$,
    
    \item $(x_{b},x_{a})\in \cR \implies ( U_{\delta_b}(x_b)\subseteq U_{\delta_a}(x_a) \wedge \forall u \in U_{\delta_b}(x_b) . \
     (x_a,u,x_a')\in\delta_a \implies \exists x_b'\in\cX_b \ . \ (x_b,u,x_b')\in\delta_b \wedge (x_b',x_a')\in\cR)$. 
\end{itemize}
\end{definition} 

%
Observe that, both definitions require that at every step the inputs in the two systems must match. 

\subsection{Strongest Asynchronous $\ell$-complete Abstractions}\label{subsec:abstraction}

In \cite{schmuck2014asynchronous,schmuck2015comparing} the authors introduce a particular class of abstractions known as SA$\ell$CA, here adapted and reformulated as a TS. One of the advantages of the SA$\ell$CA is that for its construction we only require the knowledge of the external behaviors, i.e. without the knowledge of the internal mechanisms of the underlying model, motivating our interest in this specific class when combined with data-driven techniques. 
\begin{definition}
\label{def:sl-ca}
\textnormal{\textbf{(SA$\ell$CA \cite{schmuck2015comparing})}} Let $S  \doteq (\cX,\cX_0,\cU,\delta,\cY,\cH)$ be a TS satisfying Assumption \ref{ass:concrete-sys}, and consider $\Pi_{\ell}$ and $\Pi_{\ell+1}$.
The TS
$S_\ell  \doteq  (\cX_\ell,\cX_{\ell0},\cU,\delta_\ell,\cY,\cH_\ell)$
is the SA$\ell$CA of $S$, 
where $
    \cX_{\ell} \doteq \Pi_{\ell}$ is the state set,
    $\cX_{\ell0} \doteq \{\zeta\in\Pi_{\ell} : \exists \xi \in \cI_H(S) \ . \ \cH(\xi[1-\ell,0])=\zeta \}$ is the set of initial states,  $H_\ell(\zeta) \doteq  \zeta(\ell)$ is the output map, and the transition relation is given by
    \vspace{-.2cm}
    \begin{align*}
    \delta_\ell & \doteq  \{(\zeta, u , \zeta') \ : \ \zeta[1,\ell] =\zeta'[0,\ell-1] \wedge \\
    & \zeta'|^u(\ell-1) = u \nonumber 
 \wedge \zeta\cdot\zeta'[\ell-1,\ell]\in \Pi_{\ell+1}  \} .
 \end{align*}
 \vspace{-.3cm}

\end{definition}

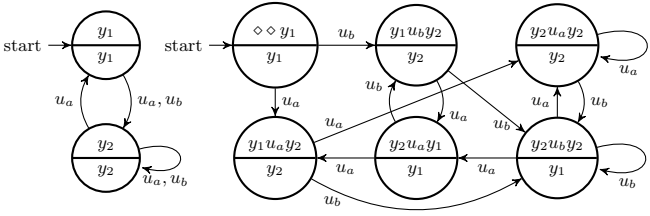
\begin{figure}[]
     \centering
     \input{models/salca-example}
     \vspace{-0.5cm}
     \caption{Illustration of the SA$\ell$CA for the system described in Example \ref{example:TS} for $\ell=0$, $S_0$ (left) and $\ell=1$, $S_1$, (right), derived using the set $\Pi_{1,4}$ and $\Pi_{2,4}$ respectively.}
     \label{fig:salca-example}
 \end{figure}

\vspace{-.5cm}
The state set of the SA$\ell$CA consists of $\ell$-sequences of the system $S$. The transition relation $\delta_\ell$ specifies a transition between two $\ell$-sequences $\zeta = y_0u_0 \ldots u_{\ell-1}y_{\ell}$ and $\zeta' = y_0'u_0' \ldots u_{\ell-1}'y_{\ell}'$ with input $u$ if the suffix of $\zeta$, i.e., $y_1u_1 \ldots u_{\ell-1}y_{\ell}$, matches the prefix of $\zeta'$, i.e., $y_0'u_0' \ldots u_{\ell-2}'y_{\ell-1}'$, $u_{\ell-1}' = u$, and if the concatenation $y_0u_0 \ldots y_{\ell}'$ belongs to $\Pi_{\ell+1}$, meaning it is an $\ell+1$-long subsequence of some $H$-long external behavior $\gamma$. This rule is known as the \emph{domino rule}.
Since the SA$\ell$CA overapproximates $S$, the value of $\ell$ determines the precision of the abstraction: larger $\ell$ values lead to tighter approximations, formally, $\cB(S) \subseteq \cB(S_{\ell+1}) \subseteq \cB(S_\ell)$. Note that $\Pi_\ell$ can be easily derived from $\Pi_{\ell+1}$.
\newline
Referring to Example \ref{example:TS}, the corresponding SA$\ell$CAs with $\ell=0$ and $\ell=1$ is shown in Fig. \ref{fig:salca-example}. 
\newline
The choice of abstracting the system as a SA$\ell$CA is motivated by the following properties:
    $(i.)$
    Knowledge of $\Pi_{\ell+1}$, derived from all external behaviors of $S$, is sufficient to construct the SA$\ell$CA.
    $(ii.)$ The set of CES can be defined based on either the \emph{past} or the \emph{future} of a state; defining CES based on the \emph{future} external behaviors is another valid approach, see \cite{schmuck2015comparing}.
    $(iii)$ When CES is defined based on the \emph{past}, as in \eqref{eq:cor-ext-beh},
    the relation
    \vspace{-.2cm}
    \begin{equation}\label{def:salca-sim-rel}
        \cR = \{(x, \zeta)\in \cX\times \cX_\ell : \ \zeta\in \cE_\ell(x)\}
    \end{equation}
    \vspace{-.6cm}
    \newline
    is a SR (w.r.t. $\cU\times\cY$) from $S$ to $S_\ell$ and the inverse relation $\cR^{-1}$ is an ASR from $S_\ell$ to $S$ provided $S$ has free input. This last claim is suggested in  \cite[Sec. V.D]{schmuck2015comparing} but not formally proven. For completeness we formalize the claim in the next proposition, proved in the Appendix.
    \vspace{-.17cm}
    \begin{proposition}\label{prop:inverse-rel-is-alt-sim}
         Consider a system $S$, let $S_\ell$ be its SA$\ell$CA as per Definition \ref{def:sl-ca} and let $\cR$ be the relation defined in \eqref{def:salca-sim-rel}. Then $\cR$ is SR from $S$ to $S_\ell$ w.r.t. $\cU\times\cY$. Further, if $S$ has free input, $\cR^{-1}$ is an ASR from $S_\ell$ to $S$ w.r.t. $\cU\times\cY$.
    \end{proposition}
    In Section \ref{sec:data-driven}, we demonstrate how to construct a data-driven version of relation \eqref{def:salca-sim-rel} and establish a similar connection with the inverse relation.
    We conclude by noting that defining CES based on the past means that $\cR$ does not need to be known explicitly. Once the SA$\ell$CA is constructed, if $(x, \zeta) \in \cR$ and applying input $u$ causes the system to transition from $x$ to $x'$ with observation $\cH(x')$, it automatically follows that there is a transition in $S_\ell$ from $\zeta$ to $\zeta'$ where $\zeta'[\ell-1, \ell] \doteq \zeta(\ell) u \cH(x')$ and $(x', \zeta') \in \cR$.


\section{Data-driven Abstractions}
\label{sec:data-driven}
Constructing a SA$\ell$CA typically requires knowledge of all possible external behaviors of the system, which can be costly in practice. To address this, we aim to build an abstraction using only sampled external behaviors.
\newline
Given a system $S_a$, we define 
the random variable $i_{\mathbf{u}_H}^a:\cX_{a0} \to \wp(\cI_H(S_a))$
\vspace{-.1cm}
\begin{equation}\label{eq:full-beh-rv}
    i_{\mathbf{u}_H}^a  \doteq \cI_H(S_a, x_{a0}, \mathbf{u}_H).
\end{equation}
\vspace{-.5cm}
\newline
Equation \eqref{eq:full-beh-rv} defines the set of full behaviors of $S_a$ for a given $\mathbf{u}_H$, starting from the randomly chosen initial condition $x_0^a$. For deterministic systems, $i_{\mathbf{u}_H}^a$ is a singleton, while for nondeterministic systems, it may be a set of behaviors.
Given a system $S_b$ and a relation $\cR\subseteq\cX_a\times\cX_b$ we define the random variable $i^b_{\mathbf{u}_H}:\cX_{a0}\rightarrow \wp(\cI_H(S_b))$
\vspace{-.2cm}
\begin{equation}\label{eq:full-beh-abs-rv}
    i_{\mathbf{u}_H}^b   \doteq \bigcup_{x_{b0}\in\cR_{\cX_b}(x_{a0})\cap\cX_{b0}}\cI_H(S_b,x_{b0},\mathbf{u}_H),
\end{equation}
\vspace{-.3cm}
\newline
which describes the set of full behaviors of $S_b$, for a given $\mathbf{u}_H$, starting from all initial conditions that are related to $x_0^a$ through $\cR$. 
\newline
Using only \emph{sampled} behaviors results in an \emph{approximation} of the SA$\ell$CA, where only a subset of $\Pi_{\ell+1}$ is available. We therefore generalize the concept of (alternating) simulation relations to account for randomly sampled initial conditions, introducing the notion of \emph{probabilistic} (alternating) simulation relations.
\vspace{-.5cm}
\newline
\begin{definition}\label{def:probabilistic-sim-rel}
\textnormal{\textbf{(Probabilistic simulation relation (PSR))}} Consider two non-blocking systems $S_a$ and $S_b$ with $\cY_a=\cY_b$ and $\cU_a=\cU_b=\cU$, and a relation $\cR\subseteq\cX_a\times\cX_b$. 
Given the probability space $(\cX_{a0},\cG_a,\mu_{x_a})$, for a fixed $\mathbf{u}_H\in\cU^H$, and with $i^a_{\mathbf{u}_H}$ and $i^b_{\mathbf{u}_H}$ defined as in \eqref{eq:full-beh-rv} and \eqref{eq:full-beh-abs-rv}, $\cR$ is a probabilistic simulation relation from $S_a$ to $S_b$ with respect to $\cU\times \cY$ until horizon $H$ with probability not less than $1-\epsilon$ if
\vspace{-.2cm}
\begin{align}\label{eq:sim-vio-easy}
    & \mu_{x_a}(x_{a0}\in \cV({S_b},\cR, H)) \leq \epsilon,
\end{align}
\vspace{-.5cm}
\newline
where the violation set $\cV({S_b},\cR, H)$ is defined 
\vspace{-.2cm}
\begin{align}    
    &\cV({S_b},\cR, H) \doteq  \{x_{a0}:\exists \mathbf{u}_H\in\cU^H \ . \ i_{\mathbf{u}_H}^a\neq\emptyset \ \wedge \nonumber \\
    &\exists \xi_a \in i_{\mathbf{u}_H}^a \ . \  (\nexists \xi_b \in i_{\mathbf{u}_H}^b \ . \ \cH_a(\xi_a)=\cH_b(\xi_b)
    \wedge \nonumber \\
    &\forall k\geq 0 \ . \ (\xi_a(k),\xi_b(k))\in\cR)\}. \label{eq:sim-vio-set}
\end{align}
\vspace{-.5cm}
\newline
More compactly, we write $\mu_{x_a}(S_a \ {}^H{\preceq}_{S.}^{\cR} S_b) > 1- \epsilon$, where ${}^H{\preceq}_{S.}^{\cR}$ highlights that the probabilistic simulation relation is referring to the relation $\cR$ and to a time horizon $H$. 
 
\end{definition}

\vspace{-.1cm}
\begin{definition}\label{def:probabilistic-alt-sim-rel}
\textnormal{\textbf{(Probabilistic alternating simulation relation (PASR))}} Under the same conditions of Definition \ref{def:probabilistic-sim-rel}, 
we say that $\cZ\subseteq \cX_b\times\cX_a$ is a probabilistic alternating simulation relation from $S_b$ to $S_a$ with respect to $\cU\times\cY$ until horizon $H$ with probability greater than $1-\epsilon$ if
\vspace{-.2cm}
\begin{align}
    &  \mu_{x_a}(x_{a0}\in \cQ({S_b},\cZ, H)) \leq \epsilon, \label{eq:alt-sim-vio-easy}
\end{align}
where the violation set $\cQ({S_b},\cZ,H)$ is defined 
\vspace{-.2cm}
\begin{align}
    &\cQ({S_b},\cZ,H) \doteq  \{x_{a0}:\exists \mathbf{u}_H\in\cU^H \ . \ i_{\mathbf{u}_H}^b\neq\emptyset \wedge (i^a_{\mathbf{u}_H}=\emptyset \vee \nonumber
    \\
    &\exists \xi_a\in i_{\mathbf{u}_H}^a \ . \  
    (\nexists\xi_b \in i_{\mathbf{u}_H}^b \ . \ \cH_a(\xi_a)=\cH_b(\xi_b)
    \wedge \nonumber
    \\
    & \forall k\geq 0 \ . \ (\xi_a(k),\xi_b(k))\in\cZ)\}. \label{eq:alt-sim-vio-set}
\end{align}
\vspace{-.6cm}
\newline
where $i_{\mathbf{u}_H}^b$ is defined as per \eqref{eq:full-beh-abs-rv} considering the relation $\cZ^{-1}\subseteq \cX_a\times \cX_b$. More compactly, we write $
    \mu_{x_a}(S_b \ {}^H{\preceq}_{A.S.}^{\cZ} S_a) > 1- \epsilon$, where ${}^H{\preceq}_{A.S.}^{\cZ}$ highlights that the probabilistic alternating simulation relation is referring to the relation $\cZ$ and to a time horizon $H$. 
 
\end{definition}
\vspace{-.1cm}
Expressions \eqref{eq:sim-vio-easy}-\eqref{eq:sim-vio-set} bound the probability of drawing an initial condition $x_{a0}$ such that there exists an input sequence $\mathbf{u}_H$, \emph{admissible for $S_a$}, which generates at least one full $H$-behavior $\xi_a$ in $S_a$ that cannot be related to any $\xi_b$ in $S_b$ by $\cR$.
Expressions \eqref{eq:alt-sim-vio-easy}-\eqref{eq:alt-sim-vio-set} bound the probability of drawing an initial condition $x_{a0}$ such that there exists an input sequence $\mathbf{u}_H$, \emph{admissible for $S_b$}, which either generates at least one full $H$-behavior $\xi_a$ that can't be related to any $\xi_b$ by $\cZ$, or no full $H$-behavior at all, if $\mathbf{u}_H$ is inadmissible for $S_a$. 
The key difference is in the validity of the input sequence: in the first case, $\mathbf{u}_H(k)$ must belong to the set $U_{a}(\xi_a(k))$, ensuring $i^a_{\mathbf{u}_H}\neq\emptyset$, in the second case it belongs to $U_{b}(\xi_b(k))$, ensuring $i^b_{\mathbf{u}_H}\neq\emptyset$. The condition $i^a_{\mathbf{u}_H}=\emptyset$ in \eqref{eq:alt-sim-vio-set} captures initial conditions where $\mathbf{u}_H$ generates behaviors in $i^b_{\mathbf{u}_H}$ but not in $i^a_{\mathbf{u}_H}$, that is $\mathbf{u}_H$ is inadmissible for $S_a$ starting at $x_{a0}$. 
\vspace{-.2cm}
\begin{remark}
    If the violation sets \eqref{eq:sim-vio-set} and \eqref{eq:alt-sim-vio-set} are empty for every choice of $H$, then $\epsilon=0$, and Definitions \ref{def:sim-rel} and  \ref{def:alt-sim-rel} are equivalent to Definitions \ref{def:probabilistic-sim-rel}  and \ref{def:probabilistic-alt-sim-rel}, respectively. In this scenario, the difference between the two pairs of definitions is purely notational: the first two use transition-based requirements, while the latter two employ trajectory-based requirements, as in \cite[Lem. 1]{alur1998alternating},  which are often easier to handle in finite horizon settings.
\end{remark}

\subsection{Constructing the Data-driven Abstraction}
\label{subsec:sampling}
%
%

%
Consider the probability spaces $(\cX_0,\cG,\mu_x)$, $(\cU^H,\mathcal{F},\mu_{\mathbf{u}_H})$: when sampling the system $S_\Sigma$ for $x\in\cX_0$ and $\mathbf{u}_H\in \cU^H$ we assume to have access only to the external $H$-behavior of the system, that is $\cB_H(S_{\Sigma},x,\mathbf{u}_H)$: note that by Assumption \ref{ass:concrete-sys}, the latter is necessarily a singleton. 
In order to construct a data-driven SA$\ell$CA of $S_{\Sigma}$ we pursue a \emph{random exploration} of the system's external behaviors, using random initial conditions and input sequences. In other words, we draw $N$ i.i.d. pairs $(x^i,\mathbf{u}_H^i)$ according to the product probability measure $\mu_p$, and we obtain the set of sampled external behaviors 
\vspace{-.2cm}
\begin{equation}\label{eq:sampled-behaviors}
    D \doteq \{\cB_H(S_{\Sigma},x^i,\mathbf{u}_H^i) \ : \ i=1,2,...N\}.
\end{equation}
\vspace{-.5cm}
\newline
For $\ell < H$ we denote by $\hat{\Pi}_{\ell}$ the set of all witnessed subsequences of length $\ell$, that is
\vspace{-.2cm}
\begin{equation}\label{eq:sampled-l-subseq}
    \hat{\Pi}_{\ell} \doteq  \bigcup_{\gamma\in D}\bigcup_{k\in[0,H]}\gamma[k-\ell,k].
\end{equation}
\vspace{-.5cm}
\newline
From now on we use the symbol $\hat{\cdot }$ as shown above to denote the quantities depending on the $N$ samples drawn according to $\mu_p$. 
We are now ready to define the data-driven SA$\ell$CA. 
\begin{definition}
\label{def:data-driven-slca}
\textnormal{\textbf{(Data-driven SA$\ell$CA)}} Given $\hat{\Pi}_{\ell+1}$, the TS  
$\hat{S}_\ell = (\hat{\cX}_\ell,\hat{\cX}_{\ell,0},\cU,\hat{\delta}_\ell,\cY,\cH_\ell)$ 
is called the data-driven (strongest asynchronous) $\ell$-complete abstraction (SA$\ell$CA) of $S_{\Sigma}$, where $\hat{\cX}_\ell  \doteq  \hat{\Pi}_{\ell}$ is the state set, $\hat{\cX}_{\ell,0}  \doteq  \{\zeta\in\hat{\Pi}_{\ell} : \exists \xi \in \cI_H(S) \ . \ \cH(\xi[1-\ell,0])=\zeta \}$ is the initial set, and the transition relation is given by
\vspace{-.3cm}
\begin{align*}
   \hat{\delta}_\ell & \doteq  \{(\zeta, u , \zeta') \ :  \zeta[1,\ell] =\zeta'[0,\ell-1] \wedge \nonumber 
    \\ \nonumber
    & \zeta'|^u(\ell-1) = u \nonumber 
 \wedge \zeta\cdot\zeta'[\ell-1,\ell]\in \Pi_{\ell+1}  \}.
\end{align*}
\end{definition}
\vspace{-.5cm}
Trivially, $\hat{\Pi}_{\ell+1}\subseteq\Pi_{\ell+1}$. In the following, we show how to derive a PASR from $\hat{S}_\ell$ to $S_{\Sigma}$ until horizon $H$ with probability not lower than $1-\epsilon$ up to some confidence $\beta$.
\begin{remark}
\label{rem:complexity}
Constructing the \salca \ using the set of sampled behaviors $D$ entails solving a scenario program \cite{coppola2023data}. Its complexity $s^*_N$ is the cardinality of the smallest subset of $D$  \eqref{eq:sampled-behaviors} which would result in the same set of all witnessed $(\ell+1)$-sequences $\hat{\Pi}_{\ell+1}$. To highlight the dependency of the complexity on the parameter $\ell$ from here on we denote it by $s^*_{N,\ell}$.
One may use Theorem \ref{theo:scenario-gurantees} using any upper bound of $s^*_{N,\ell}$; a close estimate of its value can be obtained using a greedy set cover algorithm.  
\end{remark}

The following lemma is a simple consequences of the definition above, and its proof is omitted.
\begin{lemma}\label{lem:a}
    Consider $S_{\Sigma}$, its data-driven SA$\ell$CA $\hat{S}_\ell$ constructed from the set $\hat{\Pi}_{\ell+1}$ and the relation
    \begin{equation}\label{def:data-driven-salca-sim-rel}
        \hat{\cR}  \doteq  \{(x, \zeta)\in \cX\times \hat{\cX}_\ell : \ \zeta\in \cE_\ell(x)\}.
    \end{equation} If $(x,\zeta)\in\hat{\cR}$ and $(x,u,x')\in\delta$ with $\cH(x')=y$ then there exists $(\zeta,u,\zeta')\in\hat{\delta}_\ell$ with $\cH(\zeta')=y$ and $(x',\zeta')\in\hat{\cR}$ if and only if $\zeta\cdot \zeta'[\ell-1,\ell]\in\hat{\Pi}_{\ell+1}$ with $\zeta'|^u(\ell-1)=u$. 
\end{lemma}

 By Lemma \ref{lem:a}, if a state trajectory of $S_\Sigma$ results in the same external behavior of a state trajectory of the SA$\ell$CA at every time step the pair given by the state of the former and the state of the latter are in $\hat{\cR}$. 
\begin{lemma}\label{lem:b}
    Consider $S_{\Sigma}$, a confidence $\beta$ and the data-driven SA$\ell$CA constructed from $\hat{\Pi}_{\ell+1}$. It holds that
    \begin{align}
        &\quad\mu_p^N(\mu_p ((x,\mathbf{u}_H) \in \cT (\hat{S}_\ell,H)) \leq \epsilon )\geq 1-\beta, \label{eq:salca-bounds} 
    \end{align}
    where the violation set is $\cT(\hat{S}_\ell,H)  \doteq  \{(x,\mathbf{u}_H) : \cB_H(S_{\Sigma},x,\mathbf{u}_H)\notin \cB_H(\hat{S}_\ell)\} $,
    and 
    $\epsilon \doteq \epsilon(s^*_{N,\ell},\beta,N)$ as defined in Theorem \ref{theo:scenario-gurantees}.
\end{lemma}

Lemma~\ref{lem:b}, detailed in the Appendix, provides an upper bound on the probability that a sampled external behavior from the concrete system does not belong to the set of external behaviors of the data-driven SA$\ell$CA. This is related to the concept of \emph{behavioral inclusion} \cite{tabuada2009verification}: if the set of all behaviors of one system includes the set of all behaviors of another, we say the first system behaviorally includes the second. If \eqref{eq:salca-bounds} holds, \cite{coppola2022data} indicates that $\hat{S}_\ell$ behaviorally includes $S_\Sigma$ until horizon $H$ with probability greater than $1-\epsilon$.
\newline
Lemma \ref{lem:b} follows from \cite[Section~V.C.]{campi2018general}, where $N$ samples are drawn from a countable alphabet, identifying unique symbols and bounding the probability of unseen symbols not among the $N$ samples. We extend this by associating each $(\ell+1)$-sequence with a unique symbol, thus each sample in \eqref{eq:sampled-behaviors} produces a \emph{set} of symbols; we then bound the probability of drawing a new behavior in $\cB_H(S_\Sigma)$ containing an uncollected $(\ell+1)$-sequence, corresponding to a behavior absent in $\cB_H(\hat{S}_\ell)$.
%
\newline
For the sample complexity of \eqref{eq:salca-bounds}, while the scenario approach for convex optimization with non-degenerate constraints has a sample complexity of $O(\epsilon^{-1}\ln(\beta^{-1}))$ \cite{care2011fast}, we consider a finite sample space ($\cB_H(S_\Sigma)$) and use recent results on degenerate problems \cite{garatti2021risk}. Here, the quantity $s^*_{N,\ell}$, observed \emph{a posteriori}, connects $N$, $\beta$, and $\epsilon$. For fixed $\beta$ and $N$, the violation probability is higher when all $N$ drawn symbols are distinct ($s^*_{N,\ell}=N$) than when they are identical ($s^*_{N,\ell}=1$) \cite[Section~V.C.]{campi2018general}.
\newline
In our case, two factors influence the complexity $s^*_{N,\ell}$. The first, is the ``intrinsic richness'' of a system's external behaviors: for two systems $S$ and $S'$ where $|\cB_H(S)|>|\cB_H(S')|$ and the probability of each external behavior is uniformly distributed, we can expect that the data-driven SA$\ell$CA of $S$ will result in a higher complexity than that of $S'$.
The second factor is $\ell$: reducing $\ell$ decreases the alphabet size and $s^*_{N,\ell}$, tightening the scenario bounds: referring to Fig. \ref{fig:salca-example}, note how the state set (and the number of transitions) of $S_0$ is smaller than that of $S_1$.  However, a smaller $\ell$ may reduce the SA$\ell$CA's precision, as it introduces potential spurious behaviors not present in the original system. For example, in Fig. \ref{fig:salca-example}, the behavior $y_1u_ay_2u_by_2u_ay_1$ is possible in $S_1$ and $S_0$ but not in the original system, while $y_1u_ay_2u_a$ is possible in $S_0$ but not in $S_1$ or the original system. Larger $\ell$ values reduce such spurious behaviors. 
\newline
Finally, we show that Lemmas \ref{lem:a} and \ref{lem:b} allow us to bound the probability measure of pairs $(x,\mathbf{u}_H)$ resulting in an internal behavior of $S_\Sigma$ that cannot be related by $\hat{\cR}$ to one of the abstraction $\hat{S}_\ell$.
\begin{proposition}\label{prop:pac-input-sim-rel-H}
    Consider $S_{\Sigma}$ , the product probability space $(\cP,\mathcal{W},\mu_p)$ of $(\cX,\cG,\mu_x)$ and $(\cU^H,\mathcal{F},\mu_{\mathbf{u}_H})$, and the data-driven SA$\ell$CA $\hat{S}_\ell$ obtained from the set $\hat{\Pi}_{\ell+1}$. Given a confidence parameter $\beta$, it holds that 
    \vspace{-.3cm}
    \begin{equation}
        \mu_p^N(\mu_p(
        (x,\mathbf{u}_H) \in \overline{\cV}(\hat{S}_\ell,\hat{\cR},H) ) \leq \epsilon ) \geq 1-\beta
    \end{equation}
    \vspace{-.4cm}
    where $\epsilon \doteq \epsilon(s^*_{N,\ell}, \beta, N)$ as defined in \eqref{eq:scenario-confidence}, and
    \begin{multline}\label{eq:partial-sim-vio-set}
        \overline{\cV}(\hat{S}_\ell,\hat{\cR},H) \doteq \{(x,\mathbf{u}_H):
    \exists \xi \in i_{\mathbf{u}_H} \ . \ 
    (\nexists\xi_\ell \in i_{\mathbf{u}_H}^\ell \ . \ \\\cH(\xi)=\cH_{\ell}(\xi_\ell)
     \wedge 
     \forall k\geq0 \ . \ (\xi(k),\xi_\ell(k))\in\hat{\cR})\}
    \end{multline}
\end{proposition}  
\vspace{-.8cm}
\begin{proof}
     First, we show that the probability of drawing a pair $(x,\mathbf{u}_H)$ resulting in an external $H$-behavior not contained in the set of all external $H$-behaviors of $\hat{S}_\ell$ is bounded by $\epsilon$ with confidence $1-\beta$. Consider the random variable $w_{x, \mathbf{u}_H}:\cX\times\cU^H\rightarrow\cB_H(S)$ defined as $w(x, \mathbf{u}_H) \doteq \cB_H(S,x,\mathbf{u}_H)$. From Lemma \ref{lem:b},
        $   
        \mu_p^N(\mu_p((x,\mathbf{u}_H) \ : 
        w(x, \mathbf{u}_H) \notin \cB_H(\hat{S}_\ell) ) \leq \epsilon ) \geq 1-\beta.$ 
    Suppose that $(x,\mathbf{u}_H)\in \overline{\cV}(\hat{S}_\ell,\hat{\cR},H)$
    : by Lemma \ref{lem:a}, this implies that 
    $
    w(x, \mathbf{u}_H) \notin \cB_H(\hat{S}_\ell)
    $.
    Since the negation of the latter holds with probability greater than $1-\epsilon$ then the negation of the former holds with at least the same probability, up to a confidence of at least $1-\beta$. 
\end{proof} 
\vspace{-.4cm}
The distinction between the sets $\overline{\cV}(\hat{S}_\ell,\hat{\cR},H)$ and $\cV(\hat{S}_\ell,\hat{\cR},H)$, as defined in \eqref{eq:sim-vio-set} (considering $S_b = \hat{S}_\ell$) is subtle but important. The first set includes \emph{pairs} $(x,\mathbf{u}_H)$ that lead to an external behavior in the concrete system absent in the data-driven SA$\ell$CA. The second set includes \emph{states} $x$ for which there \emph{exists} an input sequence $\mathbf{u}_H$ causing an external behavior in the concrete system not present in the data-driven SA$\ell$CA.
However, for the guarantees in Proposition \ref{prop:pac-input-sim-rel-H} to hold, pairs $(x,\mathbf{u}_H)$ must be drawn according to the product measure $\mu_p$, meaning both the initial condition $x$ and the input sequence $\mathbf{u}_H$ must be randomly sampled. Since our goal is to create an abstraction suitable for control, we need the flexibility to select inputs arbitrarily after constructing the data-driven SA$\ell$CA, rather than being constrained by the probability distribution $\mu_{\mathbf{u}_H}$.
We expand the result of Proposition \ref{prop:pac-input-sim-rel-H} to cover arbitrarily chosen control sequences.
\vspace{-.2cm}
\begin{proposition}\label{prop:pac-sim-rel}
    Consider $S_\Sigma$, the product probability space $(\cP,\mathcal{W},\mu_p)$ of $(\cX,\cG,\mu_x)$ and $(\cU^H,\mathcal{F},\mu_{\mathbf{u}_H})$, and the data-driven SA$\ell$CA $\hat{S}_\ell$ obtained from the set $\hat{\Pi}_{\ell+1}$. If $\mu_{\mathbf{u}_H}$ is uniformly distributed, given a confidence $\beta$, with $\overline{\epsilon}=\min(1,\epsilon|\cU^H|)$ it holds that 
    \vspace{-.2cm}
    \begin{equation}
         \mu_p^N(\mu_x(S_\Sigma \ {}^H{\preceq}_{S.}^{\hat{\cR}} \hat{S}_\ell) > 1- \overline{\epsilon} ) \geq 1-\beta.
    \end{equation}
\end{proposition}
\vspace{-.8cm}
\begin{proof}
For brevity, let $\cV$ and $\overline{\cV}$ represent the sets $\cV(\hat{S}_\ell,\hat{\cR},H)$, as defined in $\eqref{eq:sim-vio-set}$, and $\overline{\cV}(\hat{S}_\ell,\hat{\cR},H)$, as defined in \eqref{eq:partial-sim-vio-set}. In Proposition \ref{prop:pac-input-sim-rel-H} we have shown that $\mu_p^N(\mu_p((x,\mathbf{u}_H) \in \overline{\cV}) \leq \epsilon ) \geq 1-\beta.$
    We define the set $J(x) \doteq \{\mathbf{u}_H\in\cU^H : (x,\mathbf{u}_H)\in\overline{\cV}\}.$
    Let $c$ and $z$ be the densities of $\mu_x$ and $\mu_{\mathbf{u}_H}$ respectively. Then, 
    \vspace{-.2cm}
    \begin{align}
        &\mu_p( 
        (x,\mathbf{u}_H) \in\overline{\cV}) \label{eq:pac-sim-rel-1} = \int_{\cV}c(x)\int_{J(x)}z(\mathbf{u}_H)d\mathbf{u}_Hdx 
        \\
        &= \int_{\cV}c(x)\frac{|J(x)|}{|\cU^H|}dx
        \geq \int_{\cV}\frac{c(x)}{|\cU^H|}dx = \frac{\mu_x( 
        x \in \cV )}{|\cU^H|}, \label{eq:pac-sim-rel-3}
    \end{align}
    \vspace{-.3cm}
    \newline
    from which follows the thesis
        $\mu_x(
        x \in \cV ) \leq \epsilon |\cU^H|$,
    holding with a probability of at least $1-\beta$. 
\end{proof}
\vspace{-.5cm}
Proposition \ref{prop:pac-sim-rel} states that the probability of drawing an initial condition in $\cV$—where there \emph{exists} an input $\mathbf{u}_H$ that generates an external behavior not related by $\hat{\cR}$ to one in the SA$\ell$CA—is bounded, up to a given confidence. In other words, after sampling $N$ independent $H$-long external behaviors according to $\mu_p = \mu_x\times\mu_{\mathbf{u}_H}$ to construct $\hat{S}_\ell$, Proposition \ref{prop:pac-sim-rel} establishes a PSR from the concrete system to the abstraction, up to some confidence. If $x\notin\cV$, any arbitrary input sequence $\mathbf{u}_H$ will generate an external behavior that exists in the (data-driven) SA$\ell$CA; Thus, after constructing the abstraction, control actions in $S_\Sigma$ can be chosen arbitrarily, not constrained by the uniform distribution $\mu_{\mathbf{u}_H}$.  This property is crucial for the next step, where we show that control actions can be chosen using the SA$\ell$CA once a PASR from the SA$\ell$CA to the concrete system is established.
\begin{remark}
    Proposition \ref{prop:pac-sim-rel} assumes that $\mu_{\mathbf{u}_H}$ is uniform. The same reasoning applies to other measures, provided the smallest non-zero probability assigned to an element $\cU^H$ is known. However, if $\mu_{\mathbf{u}_H}$ is selectable, the bound in \eqref{eq:pac-sim-rel-3} is tightest with the uniform measure.
\end{remark}
Next, we claim that $\hat{\cZ} \doteq (\hat{\cR})^{-1}$ defines a PASR. 
\begin{corollary}\label{cor:alt-sim-horizon}
Under the assumptions of Proposition \ref{prop:pac-sim-rel}, $\hat{\cZ}$ defines a PASR from $\hat{S}_\ell$ to $S_\Sigma$ with respect to $\cU\times\cY$ until horizon $H$ with probability not less than $1-\overline{\epsilon}$, with $\overline{\epsilon}=\min(1,\epsilon|\cU^H|)$, with confidence greater than $1-\beta$, that is 
\vspace{-.3cm}
\begin{equation}
         \mu_p^N(\mu_x(\hat{S}_\ell \ {}^H{\preceq}_{A.S.}^{\hat{\cZ}} S_\Sigma) > 1- \overline{\epsilon} ) \geq 1-\beta.
    \end{equation}
\end{corollary}
\vspace{-.8cm}
\begin{proof}
    From Proposition \ref{prop:pac-sim-rel} we know with confidence $1-\beta$ that $\mu_x(x \in \cV(\hat{S}_\ell,\hat{\cR},H) ) \leq \epsilon |\cU^H|$. Hence, it is sufficient to show that $\cV(\hat{S}_\ell,\hat{\cR},H) \supseteq \cQ(\hat{S}_\ell,\hat{\cZ},H) $. Suppose that $x\in\cQ(\hat{S}_\ell,\hat{\cZ},H)$: since $S_{\Sigma}$ has free input, any $\mathbf{u}_H$ satisfying the first line of \eqref{eq:alt-sim-vio-set} is an admissible sequence of inputs for $S_{\Sigma}$, hence $i_{\mathbf{u}_H}\neq \emptyset$, which implies that $x\in \cV(\hat{S}_\ell,\hat{\cR},H)$. 
\end{proof}
\vspace{-.8cm}
Once we have established the PASR between the data-driven SA$\ell$CA and the concrete system, we can adopt classical synthesis methods to design a controller enforcing a desired specification, see e.g. \cite{tabuada2009verification}.



\section{Beyond the Sampling Horizon}
\label{sec:greater-horizons}

So far, the only assumption made on the system's dynamics is Assumption \ref{ass:concrete-sys}, which is general enough to apply to most black-box or \emph{unknown} systems. We have shown that, with a dataset of trajectories of length $H$, we can derive a PAC bound for a PASR between the abstraction and the concrete system for horizon $H$.
In this section, we extend these guarantees to horizons beyond $H$. As demonstrated in \cite{coppola2022data}, without further assumptions, nontrivial bounds for larger horizons are unattainable. Therefore, we introduce \emph{partially known} systems, adding assumptions to provide sufficient conditions for extending the horizon of guarantees. These are less general than Assumption \ref{ass:concrete-sys} but still fairly broad.
\newline
Consider the implications of $\cV(\hat{S}_\ell,\hat{\cR},H) \neq\emptyset$. Recall that $\hat{\cR}\subseteq\cX\times\hat{\cX}_\ell\subseteq\cX\times\cX_\ell$. If $\hat{\Pi}_{\ell+1} = \Pi_{\ell+1}$, then, as stated in Section \ref{subsec:abstraction}, $\hat{\cX}_\ell = \cX_\ell$, $\hat{\cR}=\cR$ and $\cV=\emptyset$, since the data-driven SA$\ell$CA coincides with the (complete) SA$\ell$CA. The converse also holds. The set $\cV$ can be alternatively expressed as the union of equivalence classes of the \emph{missing} $\ell+1$-sequences, i.e., those in $\Pi_{\ell+1}\setminus\hat{\Pi}_{\ell+1}$:
\begin{align}
    &\cV(\hat{S}_\ell,\hat{\cR},H) = \{x_0\in\cX : \exists \mathbf{u}_H\in\cU^H, \exists\xi\in\nonumber
    \\ 
    &\qquad\qquad\qquad\cI_H(S,x,\mathbf{u}_H),
    \exists k\geq0 \ . \ \xi'(k)\in \cK \} \label{eq:violation-as-union-of-pre}, \\
    & \qquad\qquad\cK  \doteq  \bigcup_{\zeta \in \Pi_{\ell+1}\setminus\hat{\Pi}_{\ell+1}} [\zeta]. \label{eq:missing-sequences-eq-class}
\end{align}
\newline
We have shown that the probability measure $\mu_x$ assigns to the set of initial conditions that can visit an equivalence class $[w_{\ell+1}]\notin\hat{\Pi}_{\ell+1}$—those not captured during sampling—is bounded by $\epsilon|\cU^H|$.
Given that the data-driven SA$\ell$CA was constructed using a set of sampled $H$-long external behaviors $D$, we now examine how this bound changes when considering the probability of visiting an $\ell$-sequence not acquired during sampling over a horizon $H+T$, with $T\in\naturals$. To do so, we analyze the change in measure under the system's time-reversed dynamics, leading to the lemma detailed in Appendix \ref{proof:bounded-image}.
\begin{lemma}\label{le:bounded-image-measure}
Given a compact set $\cX\subset \mathbb{R}^n$, the $p$-norm $||\cdot||_p : \cX\rightarrow\mathbb{R}$, and a function $g:\cX\rightarrow\cX$ such that  $||g(x)-g(x')||_p\leq L|| x-x'||_p$, if for constants $c,q>0$ it holds that $q^{-1}||x-x'||_p\leq||x-x'||_2\leq c ||x-x'||_p$ for all  $x,x'\in\cX$, then for any Riemann integrable set $\Sigma\subseteq\cX$ it holds that $\int_{g(\Sigma)} dV \leq (cLq)^n\int_\Sigma dV$. 
\end{lemma}
\begin{assumption}\label{ass:lip-inv}
    The map $f$ 
     in \eqref{eq:deterministic-sys} is Lipschitz invertible, where the distance is induced by the $p$-norm on $\cX$.
\end{assumption}
\begin{proposition}\label{prop:arbitrary-finite-time}
     Let $\mu_x$ and $\mu_{\mathbf{u}_H}$ be the uniform probability measures on $(\cX,\cG)$  and $(\cU^H,\cF)$ respectively. Consider $S_\Sigma$ with $\Sigma$ satisfying Assumption \ref{ass:lip-inv}, and the data-driven SA$\ell$CA $\hat{S}_\ell$ obtained from $\hat{\Pi}_{\ell+1}$. For any positive integer $T$ it holds that
    \begin{equation}
    \label{eq:arbitrary-finite-time}
    \mu_x(S_\Sigma \ {}^{T+H}{\preceq}_{S.}^{\hat{\cR}} \hat{S}_\ell) \leq \nu(\lambda)\mu_x(S_\Sigma \ {}^H{\preceq}_{S.}^{\hat{\cR}} \hat{S}_\ell)
\end{equation}
where
\vspace{-.3cm}
$$\nu(\lambda) \doteq \begin{cases}
    1 + \lambda^T\sum_{i=0}^{\tau-1}\lambda ^{-i(H+1)} \text{ for } \lambda \geq 1, \\
    \lambda^T + \sum_{i=0}^{\tau-1}\lambda ^{i(H+1)} \text{ for } 0< \lambda < 1,
\end{cases}$$ 
\vspace{-.3cm}
$$\tau = \lceil (H+T+1)/(H+1) \rceil-1,\quad \lambda = |\cU|\left(\frac{cq}{m_X}\right)^n, $$ 
and $c$, $q$ and $m_X^{-1}=L$ are defined as in Lemma \ref{le:bounded-image-measure}.
\end{proposition}
\begin{figure}[h]
  \centering
  \begin{minipage}[b]{0.48\columnwidth}
    \centering
    \includegraphics[width=\textwidth]{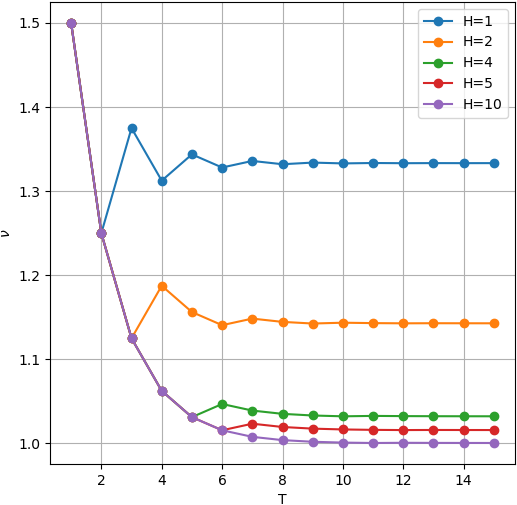}
    
    \label{fig:subfig1}
  \end{minipage}
  \hfill
  \begin{minipage}[b]{0.48\columnwidth}
    \centering
    \includegraphics[width=\textwidth]{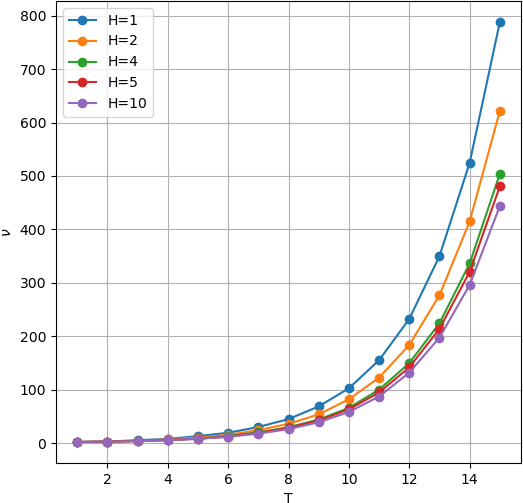}
    
    \label{fig:subfig2}
  \end{minipage}
  \caption{Plot of $\nu$ with $\lambda=0.5$ (left) and $\lambda = 1.5$ (right) for different values of $T$ and $H$.}\label{fig:nu-function}
  \label{fig:figures}
\end{figure}
Proposition \ref{prop:arbitrary-finite-time}, proved in the Appendix, connects the probability of selecting an initial condition that leads to a new behavior, absent in $\hat{S}_\ell$, within the time horizon $H$ used to construct the data-driven SA$\ell$CA, with the probability of this occurring over a horizon \emph{greater} than $H$. Note that Assumption \ref{ass:lip-inv}, and in particular the value $m_X$, is not linked to the stability of $\Sigma$ and it is satisfied by a broad set of nonlinear stable/unstable systems. Fig. \ref{fig:nu-function} shows the function $\nu$ for different parameters.


\subsection{Contracting Systems}

%
For the class of contracting systems, it is possible to show that there exists a time $H$ after which they always produce the same output. 
\begin{assumption}\label{ass:unif-contracting}
    The map $f$ of the $\Sigma$ described in \eqref{eq:deterministic-sys}, is uniformly contracting w.r.t. $x$ with constant $l_X$, uniformly Lipschitz w.r.t. $u$ with constant $l_U$, and there exist $x^*\in\cX$ and $u^*\in\mathbb{R}^{n_u}$ s.t. $f(x^*,u^*) = x^*$.
\end{assumption}
For instance, contractive control-affine systems satisfy Assumption \ref{ass:unif-contracting}  ($f(x,u)$ depends linearly on $u$ for a fixed $x$). The following lemma can be easily derived using the triangular inequality for distances.


\begin{lemma}
\label{lem:uniform-contract}
    If $f$ satisfies Assumption $\ref{ass:unif-contracting}$, then
    \begin{equation}
        d(x_k,x^*)\leq l_X^kd(x_0,x^*) + \frac{l_U}{1-l_X}\max_{0\leq i < k}\{d(u_i, u^*)\}.
    \end{equation}
\end{lemma}

\begin{proposition}\label{prop:infinite-hor}
    Let $\Sigma$ satisfying Assumption \ref{ass:unif-contracting}. If there exist a $y\in\cY$ and $r>\rho$ s.t. for all  $x\in B(x^*,r)$ it holds that $h(x_k)=y^*$ then for any $H \geq \overline{k} $ the data-driven SA$\ell$CA $\hat{S}_\ell$ of $S_{\Sigma}$ satisfies
    \begin{equation}\label{eq:measure-k-bar}
        \mu_x(S \ {}^H{\preceq}_{S.}^{\hat{\cR}} \hat{S}_\ell) = \mu_x(S \ {}^{\overline{k}}{\preceq}_{S.}^{\hat{\cR}} \hat{S}_\ell)
    \end{equation}
    where 
    \begin{equation*}
        \overline{k} = \lceil \log_{l_X} \left(r - \rho\right) - \log_{l_X}\psi\rceil,
    \end{equation*}
    \begin{equation*}
        \psi = \sup_{x\in\cX}d(x,x^*), \qquad \rho = \frac{l_U}{1-l_X}\sup_{u\in\cU}d(u,0). 
    \end{equation*}
\end{proposition} 
\vspace{-.7cm}
\begin{proof}
   After $\overline{k}$ time steps the distance of any trajectory's state $x_k$ from $x^*$ is smaller than $r$, by Corollary \ref{lem:uniform-contract} we know that $x_{i}$ will remain within that distance for all $u \in \mathcal{U}$. Moreover, by assumption of the proposition we have $h(x_k)=  y^*$ for all such $x$'s.
    Thus, the next input-output pairs are $(u, y^*)$ for all $u \in \mathcal{U}$. 
\end{proof}
\vspace{-.5cm}
The proposition assures that, if there exists a sufficiently large ball around the fixed point of $f$, where all points have the same output, the contractivity of the flow $f$ allows us to determine when all trajectories enter the ball. 
Hence, if an initial condition does not belong to the violation set defined in the right-hand side of \eqref{eq:measure-k-bar}, then neither will it belong to the violation set defined in the left-hand side.
\begin{remark}
    If the initial conditions of the system can be selected arbitrarily and not following a distribution, and if the Lipschitz constant of the system is known, other methodologies to construct an abstraction are available, see e.g. \cite{ajeleye2023data}.
\end{remark}


\subsection{Autonomous Systems}\label{subsec:autonomous-sys}

In \cite{coppola2022data}  we provide a framework for the construction of the data-driven SA$\ell$CA of an autonomous system, which, for example, can be utilized for verifying whether a linear temporal logic formula holds. As a special case, \eqref{eq:deterministic-sys} is autonomous if $|\cU|=1$. Through Proposition \ref{prop:pac-input-sim-rel-H} we obtain a guarantee that $\hat{\cR}$ is a PSR from $S_\Sigma$ to $\hat{S}_\ell$ with respect to $\cU\times \cY$ until horizon $H$ with probability not less than $1-\epsilon$ (with confidence $\beta$). Lemma \ref{lem:b} implies that $\hat{S}_\ell$ behaviorally includes $S_{\Sigma}$ until horizon $H$ with the same probability, as per \cite[Definition~4]{coppola2022data}. In summary, we recover the guarantees provided by \cite[Proposition~2]{coppola2022data}.



\section{Experimental Evaluation}
\label{sec:experiments}


\paragraph*{A Linear System}
\label{sec:experim-subsec:linear}
Let us consider the linear system $x_{k+1} = A x_k + B u_k$, where $\cU=\{ -0.3, 0, 0.3\}$,
\begin{equation}\label{eq:example}
    A = 
    \frac{1}{4}
    \begin{bmatrix}
    1 & 2 
    \\
    -1.8 & 1    
    \end{bmatrix} , 
    \quad
    B = \begin{bmatrix}
        0 \\ 1
    \end{bmatrix}.
\end{equation}
The state space $\cX = [-3, 3]^2$ is partitioned into 9 regions by a uniform grid, each uniquely labeled and defining the output set $\cY$. 
%
\newline
We sample $N=2\cdot10^6$ initial conditions $(x_0,\mathbf{u}_H)$ uniformly, with $H=4$ and $\ell=2$. 
The sampling process returns $\hat{\Pi}_{\ell+1}$, containing 342 sequences, and we construct the corresponding abstraction. 
Setting $\beta=10^{-6}$, we compute the scenario bounds according to Proposition \ref{prop:pac-input-sim-rel-H}, 
$
     \epsilon(s^*_{N,\ell}, \beta, N) = 1.51\cdot10^{-4}.
$
  Additionally, by Corollary \ref{cor:alt-sim-horizon}, with confidence at least $1-\beta$, $\hat{\cZ}$ defines a PASR from the abstraction to the concrete system with respect to $\cU\times\cY$ until horizon $H$ with probability not less than $1-
 \overline{\epsilon}$, where $\overline{\epsilon} \doteq  \epsilon(s^*_{N,\ell}, \beta, N)|\cU^H|=1.23\cdot10^{-2}$. 
In order to extend the guarantees from horizon $H=4 $ to any finite horizon we employ Proposition \ref{prop:infinite-hor}. 
%
Using the parameters $l_X \simeq 0.56$, $\psi \simeq 4.24$, $l_U = 1$, $\rho \simeq 0.68$, $r = 1$, we obtained $\overline{k} = 5$. We then construct a new abstraction after collecting $N=10^6$ trajectories with horizon $H'=5$. In line with the discussion in Section~\ref{sec:greater-horizons}, we conclude that $\hat{\cZ}$ defines a PASR from the abstraction to the concrete system until horizon $H'$, and hence any horizon, with probability not less than $1-
\overline{\epsilon}'$, where $\overline{\epsilon}'=6.18\cdot10^{-2}$, and confidence $1-\beta$. 
Alternatively, instead of resampling, we could have applied Proposition \ref{prop:arbitrary-finite-time}; we can extend the guarantee of PASR from the abstraction to the concrete system until horizon $H$ to horizon $H'$ with probability not less than $1-\nu\overline{\epsilon}=1-1.41\cdot10^{-1}$, where we have used \eqref{eq:arbitrary-finite-time} with $T=1$ to compute the correcting factor $\nu=11.4$. Proposition \ref{prop:infinite-hor} provides tighter bounds using half of the samples compared to Proposition \ref{prop:arbitrary-finite-time}.

\paragraph*{Mountain Car}

\begin{figure}[h]
  \centering
  \begin{minipage}[b]{0.48\columnwidth}
    \centering
    \includegraphics[width=\textwidth]{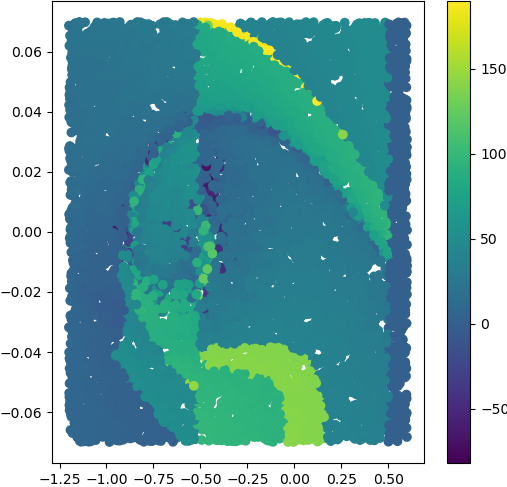}
  \end{minipage}
  \hfill
  \begin{minipage}[b]{0.48\columnwidth}
    \centering
    \includegraphics[width=\textwidth]{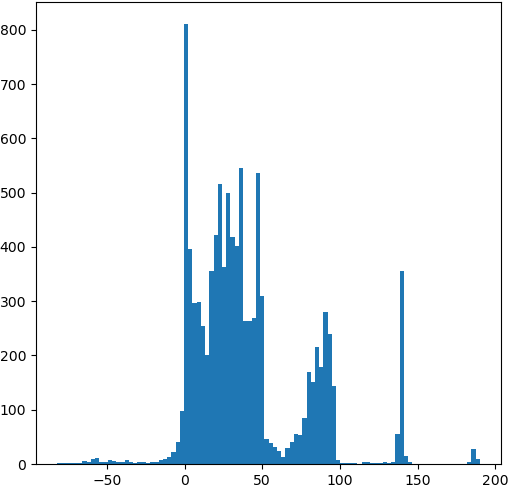}
  \end{minipage}
  \caption{The difference in time steps required to reach the goal set between controllers ($i.$) and ($ii.$) is shown on the domain $\cX$ (left) and its histogram (right), tested on $10^4$ new initial conditions.  On average the controller obtained in $(i.)$ requires 39.3 more time-steps than the one in $(ii.)$.}
  \label{fig:salca-vs-rl}
\end{figure}

\begin{figure}[h]
  \centering
  \begin{minipage}[b]{0.487\columnwidth}
    \centering
    \includegraphics[width=\textwidth]{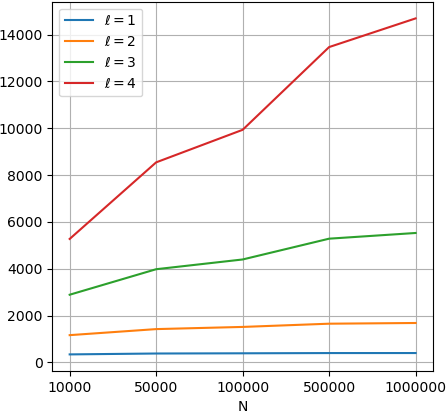}
    
    \label{fig:mc_sfig1}
  \end{minipage}
  \hfill
  \begin{minipage}[b]{0.472\columnwidth}
    \centering
    \includegraphics[width=\textwidth]{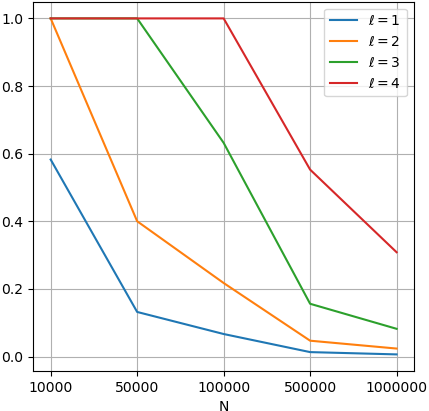}
    
    \label{fig:mc_sfig2}
  \end{minipage}
  \caption{Number of transitions in the abstraction (left) and relative $\overline{\epsilon}$ bound (right), for $\ell=1,\ldots,4$, and $\beta=10^{-3}$. }\label{fig:mountaincar}
  \label{fig:parameters}
\end{figure}

We consider and adapt the mountain car benchmark \cite{gym}.
The domain $\cX = [-1.2, 0.6] \times [-0.07, 0.07]$, uniformly sampled, accounts for position $x$ and velocity $v$. The goal of the car is to reach any point with $x\geq0.5$ (the top of a hill) as fast as possible, in at most 250 time steps. 
We compare two schemes to derive a controller with guaranteed performance, for a fixed budget of samples $N=10^6$, and confidence $\beta=10^{-3}$: ($i.$) a single-stage approach where we construct the abstraction from the uncontrolled system as per Section \ref{sec:data-driven} and derive a controller by solving a reachability game \cite{tabuada2009verification}; ($ii.$) a two-stage approach where first we design a controller using standard model-free Q-learning from reinforcement learning and then we provide performance guarantees on the controlled system, as per Section \ref{subsec:autonomous-sys}. The final result are shown in Fig. \ref{fig:salca-vs-rl}.
($i.$) We partition the domain in 6 regions, solely across the position axis, that is $[0.5, 0.6]$ is labeled $G$, and $[-1.2,0.5)$ is divided by $5$ intervals of equal length, labeled $R_1,...,R_5$.
We impose a zero-order hold control input over $T=50$ time steps, and observe the system's output accordingly every $T$ steps: this allows us to shorten the effective control horizon and improve our guarantees, at the cost of a more restrictive controller design.
We sample $N= 10^6$ pairs of initial conditions and input sequences from a uniform distribution, we collect trajectories of length $H=5$ and set $\ell=2$. Note that the system runs for $H \cdot T$ time steps in total.   
We obtain the set $\hat{\Pi}_{\ell+1}$, containing $1283$ sequences, and a complexity of $s^*_{N,\ell}=633$. This means that out of $10^6$ trajectories, 633 contain all the $\ell+1$ sequences that constitute the SA$\ell$CA. Proposition \ref{prop:pac-input-sim-rel-H} returns a bound on the violation probability for a randomly extracted pair of initial conditions and input sequences of
$
   \epsilon \doteq \epsilon(s^*_{N,\ell}, \beta, N) = 7.49\cdot10^{-4}.
$
By Proposition \ref{prop:pac-sim-rel} and Corollary \ref{cor:alt-sim-horizon}, with confidence at least $1-\beta=1-10^{-3}$, we can establish a PASR from the abstraction to the concrete system until horizon $H$ with probability not less than $1-\overline{\epsilon}$ where,
$
    \overline{\epsilon} \doteq \epsilon|\cU^H| =  2.40\cdot10^{-2}.
$
We frame the synthesis of the controller as a reachability game on the data-driven abstraction. We define as goal states all the $\ell$-sequences where the last symbol is $G$, i.e. the car is in the goal set. 
The solution of the reachability game returns a set of abstract states and actions that are guaranteed to drive the car to the goal set\footnote{By limiting the number of iterations of the reachability algorithm to $H$, we ensure that every state included in the solution reaches the goal set in no more than $H$ actions.}. 
Among the returned abstract states, there are all five $\ell$-sequences of the form $\diamond\diamond\diamond\diamond y$, where $y$ is $R_1,...,R_5$.
Hence, we can refine the controller to drive the car to the goal set from every initial state in at most $250$ time steps, with the above guarantees.
($ii.$) We use a $32\times32$ uniform grid for the domain $\cX = [-1.2, 0.5] \times [-0.07, 0.07]$, labeled $R_1,...,R_{1024}$ ($G$ as before), and define the Q-table accordingly. The learning agent receives a reward of $-1$ for every time step, until the car reaches the goal set. We allocate $N_{\text{RL}}= 5\cdot10^4$ episodes for training (exploration rate of 0.01, learning rate of 0.1) and $M = N-N_{\text{RL}}$ episodes for verifying the closed-loop SA$\ell$CA, for which we choose $\ell=100$, and obtain $\overline{\epsilon}=1.43\cdot10^{-2}$. 
\newline
\textit{Parameter study.}
We test our approach with several values of $N$, $\ell$, while maintaining a fixed time horizon of $H=5$, see Fig. \ref{fig:parameters}. We observe that, while for $\ell=1,2$ the growth of $(\ell+1)$-sequences (or transitions) rapidly tapers off with $N$, this is not the case for $\ell=3,4$; nevertheless for $N>10^{6}$ we can derive nontrivial bounds. 


\section{Discussion and Concluding Remarks}
\label{sec:discussion}
We introduced a fully data-driven framework to construct an abstraction for controlling a discrete-time deterministic system by collecting random finite length external behaviors and by leveraging a PASR. Our approach, based on the resulting data-driven SA$\ell$CA, avoids the need for reachability analysis—a common bottleneck in model-based abstractions. The number of transitions in the abstraction $\hat{\Pi}_{\ell+1}$ and consequently the number of states, adapts to the system's dynamics. The parameter $\ell$ is adjustable, allowing for more or less refined abstractions. Notably, the abstraction is agnostic to the specification, meaning if the design objective changes it can be reused to synthesize controllers without requiring resampling or reconstruction. 
\newline
Our analysis shows that the flexibility of our approach comes at the cost of the number of samples required to obtain meaningful bounds, partly because Proposition \ref{prop:pac-input-sim-rel-H} is based on the scenario approach for degenerate problems, partly because the factor $|\cU|^H$ in \eqref{eq:pac-sim-rel-3} can rapidly deteriorate the guarantees; note however that the inequality in \eqref{eq:pac-sim-rel-3} is not conservative, since it becomes an equality if each violating initial condition has exactly one input sequence $\mathbf{u}_H\in\cU^H$ that produces a new behavior. This high cost is to be expected, given that Proposition \ref{prop:pac-input-sim-rel-H} does not leverage any knowledge of the system's dynamics.
\newline
Our experiments indicate that, for a fixed budget of samples, if the control objective is known and fixed, synthesizing a controller with a technique suitable for black-box models as reinforcement learning, and successively verifying the design might result in tighter bounds, since, as argued in Section \ref{subsec:autonomous-sys}, for the closed-loop system the correcting factor of $|\cU^H|$ equals 1, and it might result in better performance. However, designing a reward function when multiple control specifications coexist (as in linear temporal logic) can be nontrivial.
\newline
%
%
We further prove that we can provide guarantees for a horizon larger than the sampling one, under additional conditions on the concrete model.
%
%
%
Future work includes the construction of abstractions for systems with noise. 
\appendix

\section{Appendix}

\textbf{Proof of Proposition \ref{prop:inverse-rel-is-alt-sim}.}
We begin by showing that $\cR$ defines a SR from $S$ to $S_\ell$ w.r.t. $\cU\times\cY$ and use this result to prove the claim of the proposition.
Pick $(x,\zeta)\in\cR$, i.e. $\zeta\in\cE(x)$, and observe that by Definition \ref{def:sl-ca} we have that $\cH_\ell(\zeta)=\zeta(\ell)$, and by \eqref{eq:cor-ext-beh} there exists an internal behavior $\xi$ of $S$ such that $\zeta=\cH(\xi[j-\ell,j])$ and $\xi(j)=x$ for some non negative integer $j$: hence $\cH_\ell(\zeta) =\cH(\xi(j))=\cH(x)$ which proves the second requirement of a SR. Next, we prove that the third requirement holds. If $(x,u,x')\in\delta$ is a transition in $S$, then there exists an internal behavior $\xi$ of $S$ such that $\zeta=\cH(\xi[j-\ell,j])$ (since $(x,\zeta)\in\cR$) and $\xi[j,j+1] = xux'$. Let $\gamma$ be the corresponding external behavior of $\xi$, i.e. $\gamma=\cH(\xi)$: by \eqref{eq:domino-tiles} we obtain that $\zeta' \doteq \cH(\xi[j-\ell+2,j+1])\in\Pi_\ell$ belongs to the state set of the SA$\ell$CA, moreover, since $\zeta\cdot\zeta'[\ell-1,\ell]= \cH(\xi[j-\ell,j+1])\in\Pi_{\ell+1}$, there exists a transition $(\zeta,u,\zeta')\in\delta_\ell$ in the SA$\ell$CA. Since $\zeta'\in\cE(x')$ we conclude that the third requirement holds. The first requirement is proved analogously. We established that $S\preceq^{\cR}_{S.}S_{\ell}$. To prove that $S_{\ell}\preceq^{\cR^{-1}}_{A.S.}S$ it is sufficient to prove the third requirement of for ASRs. To see this, observe that if $(\zeta,x)\in\cR^{-1}$ then $(x,\zeta)\in\cR$, and the third requirement of SRs implies that for every admissible input $u\in U_\delta(x)$, if $x'$ is a $u$-successor of $x$ there exists a $u$-successor $\zeta'$ of $\zeta$ such that $(x',\zeta')\in\cR$. If $S$ has free input, from the above discussion we obtain that $S_\ell$ has free input too. To conclude, $U_\delta(x)=\cU=U_{\delta_\ell}(x)$, hence the third requirement of ASRs holds. 

\textbf{Proof of Lemma \ref{lem:b}.}
\emph{Sketch.} In \cite[Prop. 1]{coppola2023data} it was shown that, for a deterministic \emph{autonomous} system $S_{\Sigma'}$, and its data-driven SA$\ell$CA $\hat{S}_\ell$ constructed from $N$ i.i.d. trajectories of length $H$ according to a probability measure $\mu$, the probability of sampling a new external behavior not existent in $\hat{S}_\ell$ can be bounded by $\epsilon$, with confidence $1-\beta$. Formally,
$
    \mu^N(\mu(\{x': \cB_H(S_{\Sigma'},x')\notin\cB_H(\hat{S}_\ell)\})<\epsilon)\geq1-\beta,
$
where $\cB_H(S_{\Sigma'},x')$ is \emph{the} external behavior generated by the autonomous $S_{\Sigma'}$ when initialized in $x'$, $\epsilon \doteq \epsilon(s^*_{N,\ell},\beta,N)$ as defined in Theorem \ref{theo:scenario-gurantees}, and $s^*_{N,\ell}$ is as per Remark \ref{rem:complexity}. Since $\mathbf{u}_H$ is available at time $0$, it is sufficient to define $\Sigma'$ on the domain $\cX\times\cU^H$ as an augmentation of $\Sigma$, where $\mathbf{u}_H$ is part of the initial conditions.

\textbf{Proof of Lemma \ref{le:bounded-image-measure}.}\label{proof:bounded-image}
The volume of a parallelotope 
$
    Q(x_1,...,x_k) = \left\{\sum_{i=1}^kr_ix_i \ : \ r_i\in[0,1]\right\}
$
is recursively computed as 
$
    \text{vol}(Q(x_1,...,x_k))  \doteq  \text{vol}(Q(x_1,...,x_{k-1}))h
$
where $h$ is the Euclidean distance of  $x_k$ from $\text{span}(x_1,...,x_{k-1})$ \cite{wu2022k}. 
Let $de_i$ be the vector of value $dx_i$ at the $i$-th component and 0 elsewhere. 
The infinitesimal element of volume shifted by $x$ is given by $x + Q(de_1,...,de_n)$ and $\text{vol}(x+Q(de_1,...,de_n)) =\text{vol}(Q(de_1,...,de_n))= \prod_{i=1}^n dx_i$. In first approximation, the shifted hypercube $x + Q(de_1,...,de_n)$ is transformed into the shifted parallelotope given by $g(x) + Q_g^n$, where $Q_g^n \doteq Q(g(x+de_1)-g(x),...,g(x+de_n)-g(x))$. We can upper bound its volume as
$
    \text{vol}(Q_g^n)\leq \text{vol}(Q_g^{n-1}) ||g(x+de_n)-g(x))||_2 
    \leq 
    \prod_{i=1}^n||g(x+de_i)-g(x))||_2. 
$
By assumption
$
    \prod_{i=1}^n||g(x+de_i)-g(x))||_2\leq \prod_{i=1}^nc||g(x+de_i)-g(x))||_p\leq
    \prod_{i=1}^ncLq||de_i||_2 = (cLq)^n\prod_{i=1}^n dx_i
$
To conclude, for the infinitesimal volume it holds that $\text{vol}(Q(g(x+de_1)-g(x),...,g(x+de_n)-g(x))) $ $\leq (cLq)^n\text{vol}(Q(de_1,...,de_n))$.


\textbf{Proof of Proposition \ref{prop:arbitrary-finite-time}.}
    For any set in $Q\subseteq\cX$ we define the following operations: $\pre{u}{Q}{} \doteq  \{x\in\cX : f(x,u)\in Q\}$, $\pre{*}{Q}{0} \doteq  Q$, and 
    \vspace{-.3cm}
    \begin{align}
        \pre{*}{Q}{k}&  \doteq  \{x\in\cX : \exists \mathbf{u}_k \in \cU^k, \nonumber \\ 
         &\xi\in\cI_H(S,x,\mathbf{u}_k) \ . \ \xi(k)\in Q \},\label{eq:pre-k} 
         \\
        \mu_x^{0,H}(Q)&   \doteq  \mu_x\left(\bigcup_{i=0}^H\pre{*}{Q}{k}\right). \label{eq:mu-pre-k}
    \end{align}
    \vspace{-.4cm}
    \newline
    By Assumption \ref{ass:lip-inv} and Lemma \ref{le:bounded-image-measure}, the pre-image of a set $Q$ is bounded, specifically
    $
        \mu_x(\pre{u}{Q}{} )\leq \lambda \mu_x(Q),
        $
    with $\lambda=\left(\frac{cu}{m_X}\right)^n$. By the union bound,
    $
        \mu_x^{0,1}(Q)\leq\bigcup_{u\in\cU}\mu_x(\pre{u}{Q}{}) +\mu_x(Q) \leq (1+\lambda)\mu_x(Q)
    $
    where $\lambda = |\cU|\eta$.  
    Note that $\pre{*}{Q}{k+1} = \pre{*}{\pre{*}{Q}{k}}{1}$. Let $E_{p}^q:=\bigcup_{i=p}^{q}\text{Pre}_*^i(Q)$. Then, for $\tau = \lceil (H + T +1)/(H+1) \rceil-1$, we can express $E_{i=0}^{H+T}$ in two equivalent forms
\begin{align}
    E_0^{H+T} = E_{j = 0}^{H} \cup \bigcup_{i = 0}^{\tau-1}E_{j = T- i(H+1)}^{H+T-i(H+1)},\label{eq:lambda-geq-1}
    \\
    E_{i=0}^{H+T} = E_{j = T}^{H+T} \cup \bigcup_{i = 0}^{\tau-1}  E_{j = i(H+1)}^{H+i(H+1)}.\label{eq:lambda-leq-1}
\end{align}
\vspace{-.3cm}
\newline
For $\lambda\geq1$, Using \eqref{eq:lambda-geq-1} and \eqref{eq:lambda-leq-1}, we derive the following bounds for $\lambda\geq1$ and $0<\lambda<1$ respectively
\begin{align}
    &\mu_x^{0,H+T}(Q) \leq \mu_x^{0,H}(Q)\left(1 + \lambda^T\sum_{i=0}^{\tau-1}\lambda ^{-i(H+1)}\right), \label{eq:mu-k-mu-k-prime-geq-1} \\
    &\mu_x^{0,H+T}(Q) \leq \mu_x^{0,H}(Q)\left(\lambda^T + \sum_{i=0}^{\tau-1}\lambda ^{i(H+1)}\right). \label{eq:mu-k-mu-k-prime-leq-1}
\end{align}
Recall \eqref{eq:missing-sequences-eq-class} and set $Q = \cK$. By combining \eqref{eq:violation-as-union-of-pre} with \eqref{eq:pre-k} and \eqref{eq:mu-pre-k} we obtain that 
$
    \mu_x(\cV(\hat{S}_\ell,\hat{\cR},H)) = \mu_x^{0,H}(\cK)
$.



\bibliographystyle{abbrv}
\bibliography{main}






\end{document}

%% file: preamble_Automatica.tex
\usepackage{algorithmic}
\usepackage{algorithm,algorithmic}
\usepackage{hyperref}
\usepackage{textcomp}
\usepackage{graphicx} 
\usepackage{subcaption}
\usepackage{amsmath,amssymb,amsfonts}
\usepackage{mathrsfs}
\usepackage{bbm}
\usepackage{scalerel}

\usepackage{tikz}
\usepackage{pgfplots}
\usetikzlibrary{automata, positioning, arrows}

\usepackage{url}
\usepackage{picins}
\usepackage{orcidlink}

\newcommand{\cU}{\mathcal{U}}

\newcommand{\cB}{\mathcal{B}} 

\newcommand{\cP}{\mathcal{P}}
\newcommand{\cX}{\mathcal{X}} 
 
\newcommand{\cG}{\mathcal{G}} 
\newcommand{\cZ}{\mathcal{Z}} 
\newcommand{\cV}{\mathcal{V}}
 
\newcommand{\cY}{\mathcal{Y}} 
\newcommand{\cF}{\mathcal{F}}
\newcommand{\cW}{\mathcal{W}}
\newcommand{\cE}{\mathcal{E}} 
\newcommand{\cH}{\mathcal{H}}
\newcommand{\cI}{\mathcal{I}}
\newcommand{\cR}{\mathcal{R}}
\newcommand{\cQ}{\mathcal{Q}}
\newcommand{\cK}{\mathcal{K}}

\newcommand{\cT}{\mathcal{T}}

\newcommand{\salca}{SA$\ell$CA}

\newcommand{\post}[2]{\text{Post}_#1(#2)}
\newcommand{\pre}[3]{\text{Pre}_{#1}^{#3}({#2})}
\newcommand{\reals}{\mathbb{R}}
\newcommand{\naturals}{\mathbb{N}}

\makeatletter
\newenvironment{definition}[1][\global\@topnum\z@]
  {\begin{defn}}
  {\end{defn}}

\newenvironment{theorem}[1][\global\@topnum\z@]
  {\begin{thm}}
  {\end{thm}}

\newenvironment{remark}[1][\global\@topnum\z@]
  {\begin{rem}}
  {\end{rem}}

\newenvironment{example}[1][\global\@topnum\z@]
  {\begin{exmp}}
  {\end{exmp}}

\newenvironment{lemma}[1][\global\@topnum\z@]
  {\begin{lem}}
  {\end{lem}}

\newenvironment{assumption}[1][\global\@topnum\z@]
  {\begin{assum}}
  {\end{assum}}

\newenvironment{proposition}[1][\global\@topnum\z@]
  {\begin{prop}}
  {\end{prop}}

\newenvironment{corollary}[1][\global\@topnum\z@]
  {\begin{cor}}
  {\end{cor}}

\newenvironment{proof}[1][\global\@topnum\z@]
  {\begin{pf}}
  {\end{pf}}
  
\makeatother

%% file: models/sm-example.tex
\tikzset{
        ->,  
        >=stealth', 
        node distance=2.5cm, 
        every state/.style={thick, minimum width=1.2cm,fill=gray!0}, 
        initial text=$ $, 
        }

\usetikzlibrary{automata, arrows.meta, positioning}

\begin{tikzpicture}[scale=0.75, transform shape, auto, initial text = start]
    
    \small

    \node (q0) [state with output, initial below] {
	$x_1$ \nodepart{lower} $y_1$
    };

    \node[state with output, right of = q0] (q1) {${x_2}$ \nodepart{lower} ${y_2}$};

    \node[state with output, left of = q0] (q2) {${x_3}$ \nodepart{lower} ${y_2}$};

    \node[state with output, right of = q1] (q3) {${x_4}$ \nodepart{lower} ${y_2}$};

    \draw (q0) edge[above] node{$u_a$} (q1);
    \draw[->, bend right] (q0) edge[below] node{$u_b$} (q2);

    \draw   (q1) edge[loop below, above] node{$u_a$} (q1);
    
    \draw[->] (q1) edge[bend right, above] node{$u_b$} (q3);

    \draw   (q2) edge[loop below, above] node{$u_b$} (q2);
    
    \draw[->, bend right] (q2) edge[above] node{$u_a$} (q0);

    \draw   (q3) edge[loop below, above] node{$u_b$} (q3);
    \draw   (q3) edge[bend right, below] node{$u_a$} (q1);



\end{tikzpicture}

%% file: models/salca-example.tex
\tikzset{
        >=stealth', 
        node distance=3cm, 
        every state/.style={thick, minimum width=1.2cm,fill=gray!0}, 
        }

\usetikzlibrary{automata, arrows.meta, positioning}
\usetikzlibrary{calc}

\begin{tikzpicture}[scale=0.75, transform shape, auto]
    
    \small
    
    \node[state with output, initial left, minimum width=1.5 cm] at (2,0) (q0) {${\diamond\diamond y_1}$ \nodepart{lower} ${y_1}$};

    \node[state with output] at (2,-2) (q3) {${y_1u_ay_2} $ \nodepart{lower} ${y_2}$};
    
    \node[state with output] at (4.5,0) (q4) {${y_1u_by_2} $ \nodepart{lower} ${y_2}$};
    
    \node[state with output] at (7,0) (q5) {${y_2u_ay_2} $ \nodepart{lower} ${y_2}$};

    \node[state with output] at (4.5,-2) (q6) {${y_2u_ay_1} $ \nodepart{lower} ${y_1}$};
    
    \node[state with output] at (7,-2) (q7) {${y_2u_by_2} $ \nodepart{lower} ${y_1}$};

    \node[state with output, initial left] at (-1,0) (q8) {${y_1}$ \nodepart{lower} ${y_1}$};

    \node[state with output] at (-1,-2) (q9) {${y_2}$ \nodepart{lower} ${y_2}$};

    \draw[->] (q0) edge[] node{$u_a$} (q3);
    \draw[->] (q0) edge[] node{$u_b$} (q4);

    \draw[->] (q3) edge[above, pos=0.1] node{$u_a$} (q5);
    \draw[->] (q3) edge[bend right, below, pos=0.1] node{$u_b$} (q7);

    \draw[->] (q4) edge[bend left, pos=0.6] node{$u_a$} (q6);
    \draw[->] (q4) edge[below, pos=0.7] node{$u_b$} (q7);

    \draw[->] (q5) edge[loop right, pos=0.8, below] node{$u_a$} (q5);
    \draw[->] (q5) edge[bend left] node{$u_b$} (q7);

    \draw[->] (q6) edge[] node{$u_a$} (q3);
    \draw[->] (q6) edge[bend left, pos=0.6] node{$u_b$} (q4);

    \draw[->] (q7) edge[loop right, pos=0.8, below] node{$u_b$} (q7);
     \draw[->] (q7) edge[] node{$u_a$} (q5);
     \draw[->] (q7) edge[] node{$u_a$} (q6);

     \draw[->] (q8) edge[bend left] node{$u_a,u_b$} (q9);
     \draw[->] (q9) edge[bend left] node{$u_a$} (q8);
     \draw[->] (q9) edge[loop right, pos=0.8, below] node{$u_a,u_b$} (q9);

\end{tikzpicture}